\documentclass{aa}

\usepackage{multirow}
\usepackage{natbib}
\usepackage{graphicx}
\usepackage{subcaption}
\usepackage{txfonts}
\usepackage{arydshln}
\usepackage[breaklinks, colorlinks, citecolor=blue]{hyperref}
\begin{document}

%%%% begin
\title{Multi-tracer analysis of straight depolarisation canals in the surroundings of the 3C~196 field\thanks{The Faraday cubes are only available in electronic form at the CDS via anonymous ftp to \url{cdsarc.u-strasbg.fr} (\url{130.79.128.5)}}}

\author{Luka Turi\'c\inst{1} \and Vibor Jeli\'c\inst{1} \and Rutger Jaspers\inst{2} \and Marijke Haverkorn\inst{2}  \and Andrea Bracco\inst{1} \and \\ Ana Erceg\inst{1} \and Lana Ceraj\inst{1} \and Cameron van Eck\inst{3} \and Saleem Zaroubi\inst{4,5,6}}

\institute{Ru{\dj}er Bo\v{s}kovi\'c Institute, Bijeni\v{c}ka cesta 54, 10000 Zagreb, Croatia\\
\email{lturic@irb.hr, vibor@irb.hr}
\and Department of Astrophysics/IMAPP, Radboud University, P.O. Box 9010,
6500 GL Nijmegen, The Netherlands
\and Dunlap Institute for Astronomy and Astrophysics, University of Toronto, 50 St. George Street, Toronto, ON M5S 3H4, Canada
\and Department of Natural Sciences, Open University of Israel, 1 University Road, P.O. Box 808, Ra’anana 4353701, Israel
\and Kapteyn Astronomical Institute, University of Groningen, P.O. Box 800, 9700AV Groningen, the Netherlands
\and Department of Physics, The Technion, Haifa 32000, Israel}

%\date{Received September 15, 1996; accepted March 16, 1997}

% \abstract{}{}{}{}{} 
% 5 {} token are mandatory
 
  \abstract
    {Faraday tomography of a field centred on the extragalactic point source 3C~196 with the LOw Frequency ARray (LOFAR) revealed an intertwined structure of diffuse polarised emission with straight depolarisation canals and tracers of the magnetised and multi-phase interstellar medium (ISM), such as dust and line emission from atomic hydrogen (HI).} 
   {This study aims at extending the multi-tracer analysis of LOFAR data to three additional fields in the surroundings of the 3C~196 field. For the first time, we study the three-dimensional structure of the LOFAR emission by determining the distance to the depolarisation canals. 
    }
   {We used the rolling Hough transform to compare the orientation of the depolarisation canals with that of the filamentary structure seen in $\text{H}\textsc{I,}$ and based on starlight and dust polarisation data, with that of the plane-of-the-sky magnetic field. Stellar parallaxes from $Gaia$ complemented the starlight polarisation with the corresponding distances.
   }
   {Faraday tomography of the three fields shows a rich network of diffuse polarised emission at Faraday depths between $-10~{\rm rad~m^{-2}}$ and $+15~{\rm rad~m^{-2}}$.  A complex system of straight depolarisation canals resembles that of the 3C~196 field. The depolarisation canals align both with the $\text{H}\textsc{I}$ filaments and with the magnetic field probed by dust. The observed alignment suggests that an ordered magnetic field organises the multiphase ISM over a large area ($\sim$20$^{\circ}$).
   In one field, two groups of stars at distances below and above 200 pc, respectively, show distinct magnetic field orientations. These are both comparable with the orientations of the depolarisation canals in the same field. We conclude that the depolarisation canals likely trace the same change in the magnetic field as probed by the stars, which corresponds to the edge of the Local Bubble.
   }
   {}

   \maketitle
%
%________________________________________________________________
\section{Introduction}
The interstellar medium (ISM) is the mass reservoir for star formation in the Galaxy. It is permeated with matter, radiation, and magnetic fields. The matter consists of a mixture of ionised, atomic, and molecular gas, dust grains, and cosmic-rays \citep[e.g.][]{mckee77,heiles12}. The process that converts interstellar matter into stars is highly multi-scale and multi-phase, governed by the interplay of thermal instability, magnetised turbulent motions, gravity, and stellar feedback \citep[e.g.][]{Hennebelle2012,Andre2014,Hennebelle2019}. In order to unveil this complex cycle of matter that regulates star formation and Galactic evolution, dedicated studies of the diffuse ISM and the dense molecular gas that hosts stellar embryos are needed. Here, we focus on the diffuse component of the ISM in the Milky Way through the unprecedented view of synchrotron emission provided by the LOw Frequency ARray \citep[LOFAR,][]{haarlem13}. 

Synchrotron emission is non-thermal radiation produced mostly by relativistic cosmic-ray electrons, and to some extent by positrons, which spiral around magnetic field lines. The emission is highly linearly polarised. The degree of its intrinsic polarisation varies between $69\%$ and $75\%$, depending on the slope of the cosmic-ray energy spectrum 
\citep[e.g.][]{rybicki86,padovani21}. 

At low radio frequencies (100 -- 200 MHz), about 70\% of the emission is expected to be intrinsically polarised, based on the observed synchrotron spectral index in total intensity \citep{guzman11,mozdzen17}, which reflects the slope of the cosmic-ray energy spectrum. However, only a few percent of the originally polarised emission is observed \citep{jelic14, jelic15, lenc16, vaneck17, vaneck19} because of depolarisation effects associated with Faraday rotation and because of the degree of regularity in the magnetic field along the sight line.

As a linearly polarised wave at a wavelength $\lambda~{\rm [m]}$ propagates through a magnetised interstellar plasma, its polarisation angle $\psi~{\rm [rad]}$ is Faraday rotated by
\begin{equation}
   \Delta \psi = \lambda^2 \Phi =\lambda^2 \left(0.81\int_0^d n_e B_{||} dl \right),
\end{equation}
where $\Phi~{\rm [rad~m^{-2}]}$ is the Faraday depth, $n_e~{\rm [cm^{-3}]}$ is the density of the thermal electrons, $B_{||}~{\rm [\mu G]}$ is the magnetic field strength of the component parallel to the line of sight $l~{\rm [pc]}$, and the integral is taken from the source ($l=0$) to the observer ($l=d$). The Faraday depth is positive when the magnetic field component parallel to the line of sight points towards the observer, while it is negative when it is in the opposite direction. For more details about the correct sense of Faraday rotation, we refer to \citet{ferriere21}.
\begin{table*}[ht!]
\caption{Overview of the observational parameters of the three LOFAR-HBA observations.}       
\label{tab:obs}      
\centering          
\begin{tabular}{l l l l}    
\hline\hline       
   Field & A & B & C \\ \hline               
   Observation ID & L428654 & L431160 & L431596 \\
   Start time [UTC] & 27-Jan-2016 20:11:00 & 10-Feb-2016 20:11:00 & 16-Feb-2016 18:37:52 \\
   End time [UTC] & 28-Jan-2016 03:51:00 & 11-Feb-2016 03:51:00 & 17-Feb-2016 02:17:52 \\
   Phase centre: RA, Dec (J2000) & $07^{\mathrm{h}}\,26^{\mathrm{m}}\,42^{\mathrm{s}}$, $+48^{\circ}\,12\arcmin\,00\arcsec$ & $08^{\mathrm{h}}\,13^{\mathrm{m}}\,30^{\mathrm{s}}$, $+40^{\circ}\,24\arcmin\,00\arcsec$ & $08^{\mathrm{h}}\,44^{\mathrm{m}}\,42^{\mathrm{s}}$, $+33^{\circ}\,54\arcmin\,00\arcsec$ \\
   Phase centre: l, b & $169.807587^{\circ}$, $+25.520618^{\circ}$ & $180.389768^{\circ}$, $+32.212486^{\circ}$ & $189.488030^{\circ}$, $+37.207857^{\circ}$ \\ \hline
   Observing frequency range & 115 -- 175\,MHz \\
   Observing spectral resolution & 3.05\,kHz \\
   Observing integration time & 2\,s \\
   Observing time & 7\,h 40\,min \\ \hline        
   \multicolumn{4}{l}{Frequency range used in Faraday tomography: 115 -- 150\,MHz with 183 kHz sub-band width}\\\hline   
\end{tabular}
\end{table*}
In the Milky Way, where the distributions of thermal and cosmic-ray electrons are mixed throughout the entire volume, differential Faraday rotation occurs and depolarises the synchrotron emission \citep{wieringa93, sokoloff98, shneider14}. As Faraday rotation is proportional to $\lambda^2$, depolarisation at low radio frequencies is more prominent than at high frequency. Nevertheless, the small amount of polarised emission that we observe carries valuable information about the physical properties of the intervening magnetised plasma.

A common technique used for analysing radio-polarimetric data is rotation measure (RM) synthesis \citep{burn66, brentjens05}. This technique decomposes the observed polarised synchrotron emission by the amount of Faraday rotation it has experienced, that is, it allows us to perform so-called Faraday tomography. It is commonly used over the whole radio band \citep[e.g.][]{jelic14, jelic15, lenc16, vaneck17, dickey19, thomson19, wolleben19, vaneck19}.

Because of the frequency dependence of Faraday rotation and the wide frequency coverage of low-frequency instruments, such as LOFAR and the Murchison Widefield Array \citep[MWA;][]{tingay13}, we can perform Faraday tomography at low-radio frequencies at a resolution in Faraday space that is a few orders of magnitude better than at GHz frequencies. At the same time, the maximum width of a structure in Faraday depth that an instrument is sensitive to, the so-called Faraday thickness \citep{brentjens05}, also depends on the frequency. A Faraday-thick structure at low radio frequencies can be Faraday thin at high frequency. This therefore needs to be taken into account when Faraday tomography results are interpreted, and the limitations in different frequency regimes need to be considered \citep{jelic15, vaneck17, vaneck18}.

LOFAR observations of the 3C 196 field showed an astonishing variety of structures of Galactic polarised synchrotron emission varying with Faraday depth \citep[hereafter Faraday structures,][]{jelic15}. The most striking features, which were seen for the first time, are very straight and long depolarisation canals. Some of them are a few degrees long and seem to extend outside the field of view. They are the result of beam depolarisation in regions of abrupt change in polarisation angle \citep{haverkorn04}, while their straightness is likely associated with the projection of the ordered and coherent magnetic field on the scale of the field of view \citep{jelic18}. The depolarisation canals as well as some Faraday structures were found to be aligned in the same region with tracers of the neutral ISM. The alignment was observed with the filamentary structure of the $21~{\rm cm}$ brightness temperature of atomic hydrogen (HI) and with the plane-of-sky magnetic field component that is probed by interstellar dust polarised emission at 353 GHz \citep{zaroubi15, kalberla17, vaneck17, jelic18, clark19, bracco20}. 

A crucial and pending issue on these correlations observed between LOFAR synchrotron polarisation and tracers of the neutral ISM is understanding their 3D structure and their relative location along the line of sight. Recent studies suggested the possible association of the observed Faraday structures with the edge of the Local Bubble \citep{jelic15,jelic18}, a 100–200 pc wide cavity around the Sun \citep[e.g.][]{lallement14}. Today, powerful diagnostics of the 3D structure of the neutral ISM are represented by accurate and sensitive measures of parallaxes \citep[e.g.][]{gaiaDR2dist} and interstellar dust extinction towards stars across the Galaxy \cite[e.g. ][]{Zucker19,Green19,Lallement19,Leike20}.

In addition to providing a precise probe of the 3D density field in the ISM, stars can also provide information about the morphology of the 3D magnetic field \citep[e.g.][]{Tassis18,Panopoulou19}. This is possible because the visible starlight is polarised by differential extinction of aspherical dust grains that are aligned with the Galactic magnetic field \citep{Hall49,Hiltner49,davis51}.

\begin{figure}[!t]
\centering \includegraphics[width=.9\linewidth]{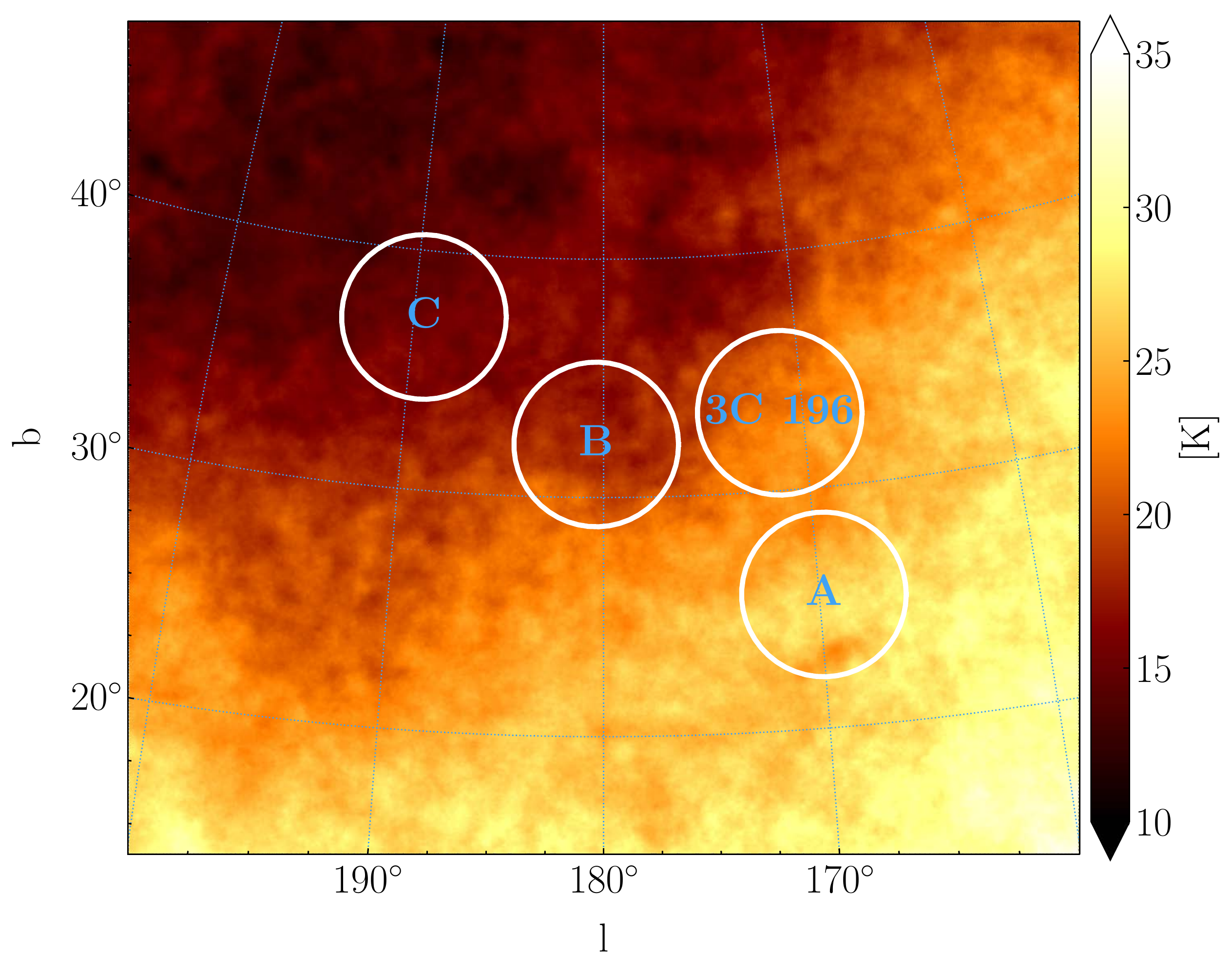}
\caption{Stereographic projection of the Haslam 408~MHz map \citep{remazeilles2015} of the area nearby the 3C 196 field, given in Galactic coordinates. Positions of the three fields studied in this work (Fields A, B, and C) as well as the position of the 3C 196 field are marked with circles. The size of each circle corresponds to the field of view of each observation ($37.6~{\rm deg^2}$). For more details, see Sec.~\ref{sec:obs} and Table~\ref{tab:obs}.}
\label{fig:Haslam_3C196area}
\end{figure}

In this paper, we have two main goals. First, we aim at extending the study of Faraday structures and depolarisation canals in LOFAR observations of three fields located nearby the 3C 196 field (see Fig.\ref{fig:Haslam_3C196area}). Field A is located $10^\circ$ below the 3C 196 field, towards the Galactic plane; Field B is at a similar Galactic latitude as the 3C 196 field, but $10^\circ$ towards higher Galactic longitudes; and Field C is in the coldest part of the north Galactic halo. Second, for the first time, we use starlight polarisation data and stellar parallaxes to investigate the spatial distribution of the observed correlation between the depolarisation canals and the neutral ISM as a function of distance. The LOFAR observations and the Faraday cubes presented in this paper were also used by \citet{bracco20} to study the morphological correlation of synchrotron polarisation with the multi-phase and magnetised HI. The authors found that most of the polarised emission seen in LOFAR data is correlated to the filamentary structure of the cold ($< 100$ K) $\text{H}\textsc{I}$ gas (hereafter, cold neutral medium, or CNM).

The paper is organised as follows. LOFAR observations and related data products are described in Sect.~2. Section~3 presents the Faraday tomography of the diffuse polarised emission. The multi-tracer analysis of the depolarisation canals observed in three fields is presented in Sect.~4, while Sect.~5 presents the results based on the starlight polarisation data. The paper concludes with summary and conclusions in Sect.~6.  

\section{Observations and data processing}\label{sec:obs}
The LOFAR High Band Antennas (HBA) observations of the three fields analysed in this paper were taken in January and February 2016 (under project code LC5\_008). The array was used in the HBA DUAL INNER configuration \citep{haarlem13}, where the remote stations were tapered to be the same as the core stations. The total observing time of each observation was 7 hours and 40 minutes, with a correlator integration time of 2 s. The observations were symmetric around transit, and they were taken during night time. The observing frequency range was between 115 MHz to 175 MHz, divided into 308 frequency sub-bands of 195.3125 kHz width. The spectral resolution was 3.05 kHz. The phase centres of each observation (see  Fig.~\ref{fig:Haslam_3C196area}), together with their observational parameters, are given in Table~\ref{tab:obs}.

The initial processing and calibration were performed following the steps described in \citet{jelic15}. Here we give a brief overview of these steps. First, we applied automatic flagging of the radio-frequency interference (RFI) with \texttt{AOFlagger} \citep{offringa10, offringa12}, after which we averaged the data in time and frequency. The resulting data have a time resolution of 12 s and a spectral resolution of 183 kHz because the first and last two 3.05 kHz channels of each frequency sub-band were removed. The averaged data were then calibrated in a direction-independent manner using \texttt{Black Board Selfcal} \citep{pandey09}. The sky-model used for the calibration was generated by the \texttt{Global Sky Model}\footnote{\url{http://github.com/bartscheers/gsm}}. It includes sources brighter than 1.5 Jy from the VLA Low-Frequency Sky Survey (VLSS) catalogue at 74 MHz \citep{cohen07} and located within $3^\circ$ from the phase centre of each observation. 

We also corrected the data for the Faraday rotation produced by the Earth's ionosphere using \texttt{RMextract} \citep{mevius18, mevius18soft}. The ionospheric Faraday rotation was about $0.3~{\rm rad~m^{-2}}$ throughout the first two observations (Fields A and B), and it gradually decreased from $0.8~{\rm rad~m^{-2}}$ to $0.3~{\rm rad~m^{-2}}$ during the third observation (Field C).

\begin{figure}[!t]
\centering \includegraphics[width=.9\linewidth]{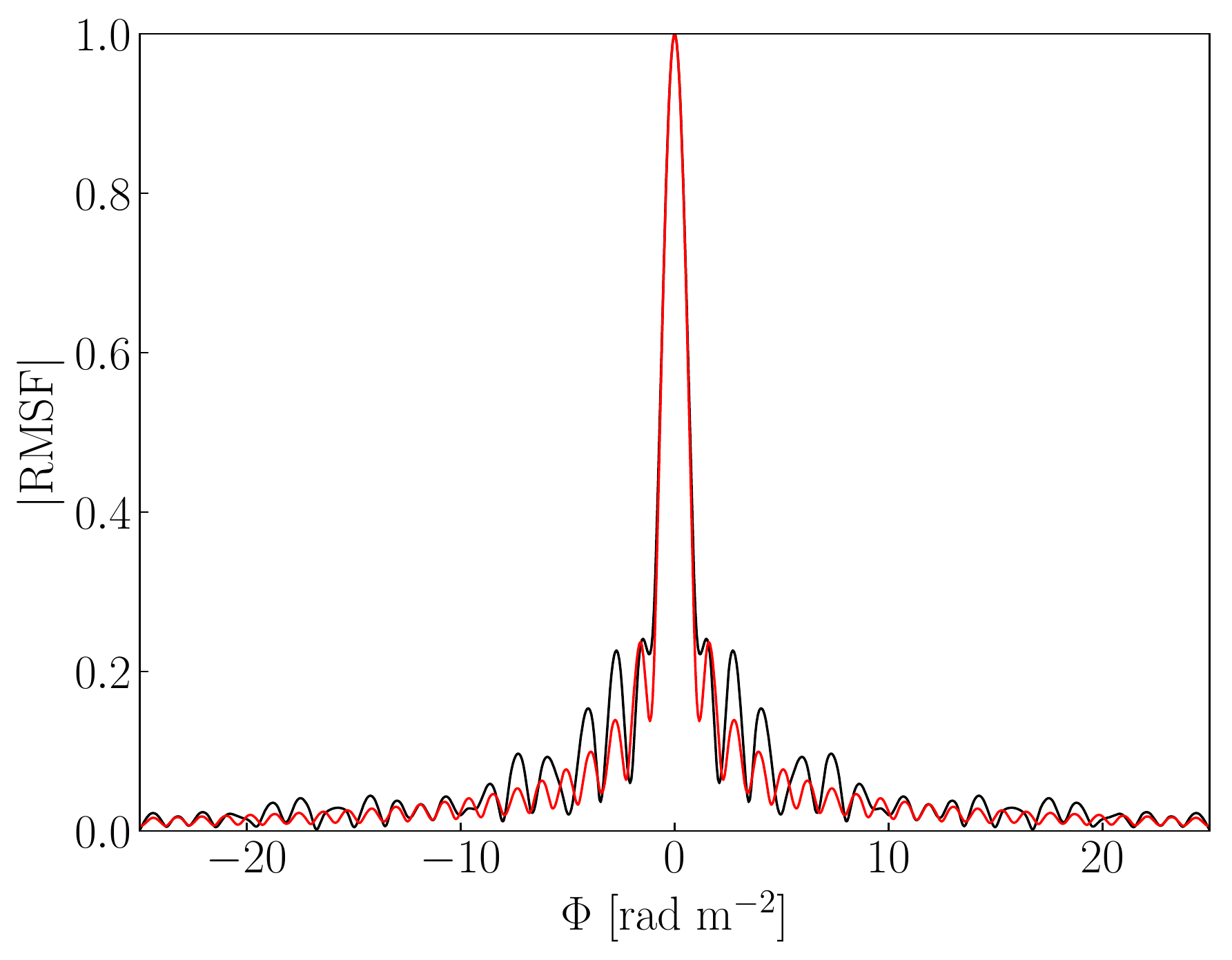}
\caption{Absolute value of the RMSF shown in black for Field A and in red for Fields B and C. The FWHM of the fitted Gaussian to the central peak of the RMSF gives the resolution in Faraday depth, $\delta\Phi=1.8~{\rm rad~m^{-2}}$.}
\label{fig:RMSF}
\end{figure}

The data were imaged in all Stokes parameters ($I$, $Q$, $U$ and $V$) using \texttt{excon} \citep{yatawatta14}. We used baselines between 10 and 800 wavelengths weighted with the Briggs scheme, using a robustness parameter equal to 0. This resulted in a frequency-independent angular resolution of $3.9\arcmin\times3.6\arcmin$, defined by the size of the point spread function (PSF). 
Finally, to perform Faraday tomography, we applied \texttt{rmsynthesis}\footnote{\url{http://github.com/brentjens/rm-synthesis}} \citep{brentjens05} to the Stokes $Q$ and $U$ images of the $170$ sub-bands, which had comparable noise levels ($\lesssim 2.5~{\rm mJy~PSF^{-1}}$). These 183 kHz sub-bands covered frequencies between $115$ MHz and $150$ MHz. The frequencies above 150 MHz were highly affected by broadband RFI. If frequencies above $150$ MHz, not affected by the RFI, were included in Faraday tomography, the sparse frequency coverage would give rise to three times higher sidelobes of the rotation measure spread function (RMSF) at $|\Phi| > 10~{\rm rad~m^{-2}}$. On the other hand, limiting our frequency coverage to the frequencies below $150$ MHz resulted in a lower resolution in Faraday depth of $1.8~{\rm rad~m^{-2}}$, as compared to $0.9~{\rm rad~m^{-2}}$ if all frequencies were taken into account. Observation of Field A was more affected by RFIs than the observations of Fields B and C, resulting in higher sidelobes (see Fig.~\ref{fig:RMSF}). The final cubes covered a Faraday depth range from $-25~{\rm rad~m^{-2}}$ to $+25~{\rm rad~m^{-2}}$ in $0.25~{\rm rad~m^{-2}}$ steps. Because the resolution was higher than the maximum detectable scale ($\Delta\Phi=0.8~{\rm rad~m^{-2}}$), we were only sensitive to Faraday-thin structures ($\lambda^2\Delta\Phi\ll1$) or the edges of Faraday-thick structures \citep[$\lambda^2\Delta\Phi\gg1$, ][]{brentjens05,vaneck17}. We did not deconvolve the Faraday cubes for the sidelobe effect of the RMSF. 

\begin{table}[t!]
\caption{Noise in the Faraday cubes of the observed fields, defined as a standard deviation of an image at $\Phi=-25~{\rm rad~m^{-2}}$. The values are given for the polarised intensity (PI) and the Stokes $Q$ and $U$.}
\label{tab:RMnoise}      
\centering                          
\begin{tabular}{c c c c}        
\hline\hline                 
\multirow{2}{*}{Field} & $PI$ & $Q$ & $U$ \\    
   & \multicolumn{3}{c}{$\mathrm{[\mu Jy\,PSF^{-1}\,RMSF^{-1}]}$} \\ \hline
   A & $108$ & $164$ & $164$ \\
   B & $110$ & $166$ & $168$ \\
   C & $89$ & $135$ & $135$ \\ \hline    
\end{tabular}
\end{table}

The noise in the Faraday cubes for the different observations is given in Table~\ref{tab:RMnoise}. It is calculated as the standard deviation of an image at $\Phi=-25~{\rm rad~m^{-2}}$, where we do not observe any polarised emission and the image is dominated by noise. The noise is comparable between Fields A and B, while in the case of Field C, it is $\sim1.2$ times lower. This difference can be attributed to the lower sky brightness in Field C than in Fields A and B (see Fig.~\ref{fig:Haslam_3C196area}). The system noise temperature, which defines the noise in the images at each frequency and consequently in the Faraday cubes, depends on the instrumental noise temperature and on the sky brightness temperature. Hence, the lower the sky brightness contribution, the lower system noise temperature. For all LOFAR frequencies, the sky brightness temperature is dominated by the frequency-dependent Galactic emission \citep{haarlem13}. The noise in the Faraday cube of polarised intensity is consistent with normally distributed noise in Stokes $Q$, $U$.
\begin{table*}[!ht]
\caption{Overview of the brightness temperature and Faraday depth of the detected polarised emission in the observed fields. We also provide the brightness temperature in total intensity at $408$~MHz \citep{remazeilles2015} and $1.4$~GHz \citep{Reich1981, Reich1986, Reich2001}, the brightness temperature of the polarised intensity at $1.4$~GHz \citep{wolleben06}, the brightness temperature of the total intensity scaled down from $408$~MHz to $130$~MHz using the spectral index from \citet{guzman11}, and the calculated polarisation fractions at $130$~MHz and $1.4$~GHz.}       
\label{tab:polfrac}      
\centering          
\begin{tabular}{l|l| l l l}    
\hline\hline       
   \multicolumn{2}{l|}{Field} & A & B & C \\ \hline               
   \multicolumn{2}{l|}{RM range $\Phi~\mathrm{[rad\,m^{-2}}]$} & $-6$ -- $+8$ & $-4$ -- $+8$ & $+5$ -- $+15$ \\
   \multicolumn{2}{l|}{Integrated polarised intensity $T_{130\,\mathrm{MHz}}~\mathrm{[K]}$} & $5 \pm 3$ & $1.5 \pm 0.9$ & $0.6 \pm 0.4$ \\ \hline    
   \multicolumn{2}{l|}{Total intensity $T_{408 \,\mathrm{MHz}}~\mathrm{[K]}$} & $26.1 \pm 0.9$ & $18.0 \pm 0.5$ & $15.4 \pm 0.5$ \\
   \multicolumn{2}{l|}{Spectral index $\beta$} & $-2.475$ & $-2.500$ & $-2.525$ \\ \hline
   \multicolumn{2}{l|}{Total intensity $T_{408\,\mathrm{MHz} \rightarrow 130\,\mathrm{MHz}}\,\mathrm{[K]}$} & $440 \pm 20$ & $314 \pm 9$ & $277 \pm 9$ \\
   \multicolumn{2}{l|}{Polarisation fraction [\%] } & $1.1 \pm 0.7$ & $0.5 \pm 0.3$ & $0.2 \pm 0.1$ \\ \hline
   \multicolumn{2}{l|}{Total intensity $T_{1.4 \,\mathrm{GHz}}~\mathrm{[K]}$} & $3.59 \pm 0.03$ & $3.43 \pm 0.03$ & $3.32 \pm 0.02$\\
   \multicolumn{2}{l|}{Polarised intensity $T_{1.4 \,\mathrm{GHz}}~\mathrm{[K]}$} & $0.17 \pm 0.01$ & $0.087 \pm 0.008$ & $0.081 \pm 0.006$\\
    \multicolumn{2}{l|}{Polarisation fraction [\%] } & $4.7 \pm 0.3$ & $2.5 \pm 0.2$ & $2.4 \pm 0.2$ \\ \hline
\end{tabular}
\end{table*}

\begin{figure*}[!ht]
\centering \includegraphics[width=.45\linewidth]{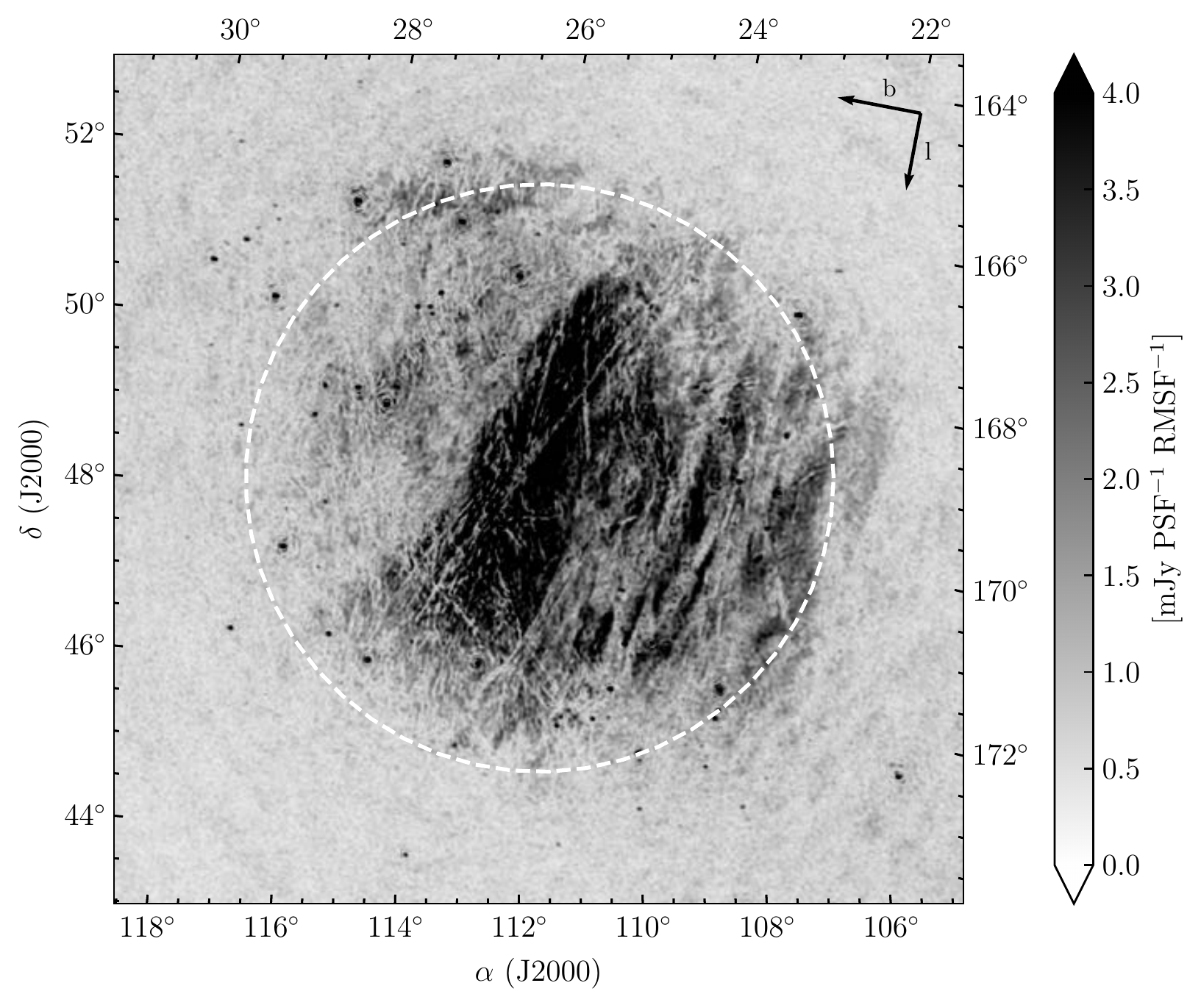}
\centering \includegraphics[width=.45\linewidth]{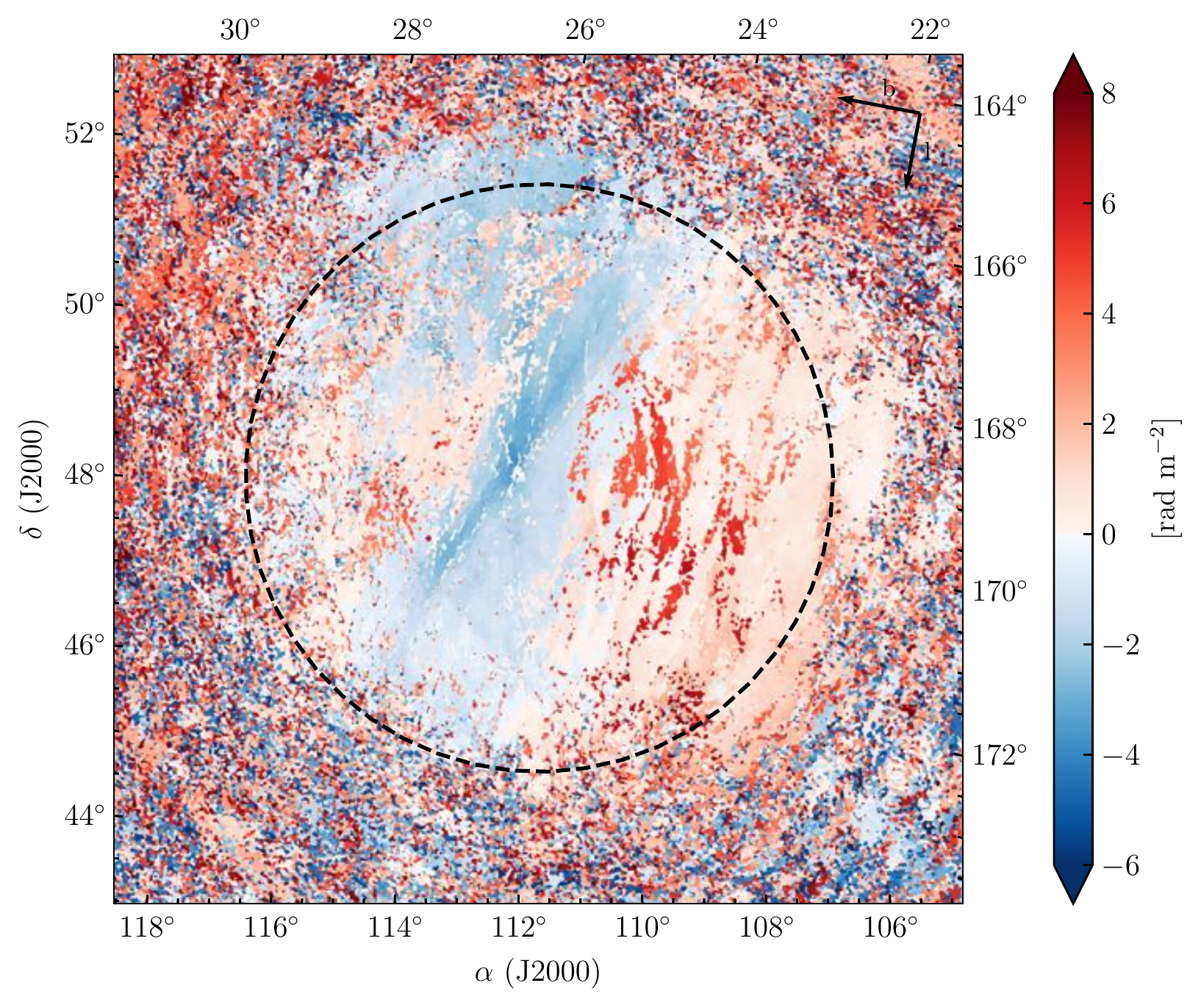}
\centering \includegraphics[width=.45\linewidth]{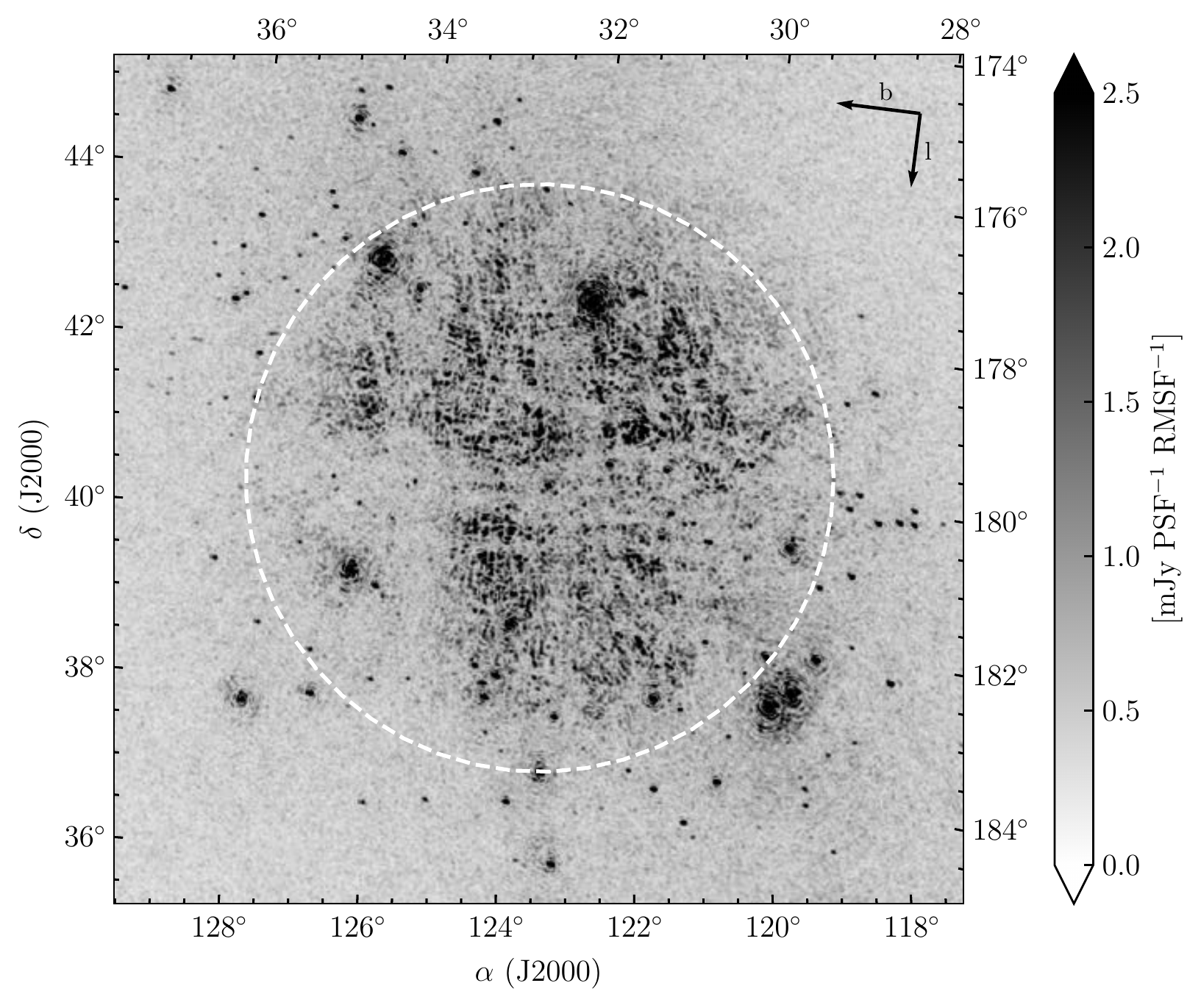}
\centering \includegraphics[width=.45\linewidth]{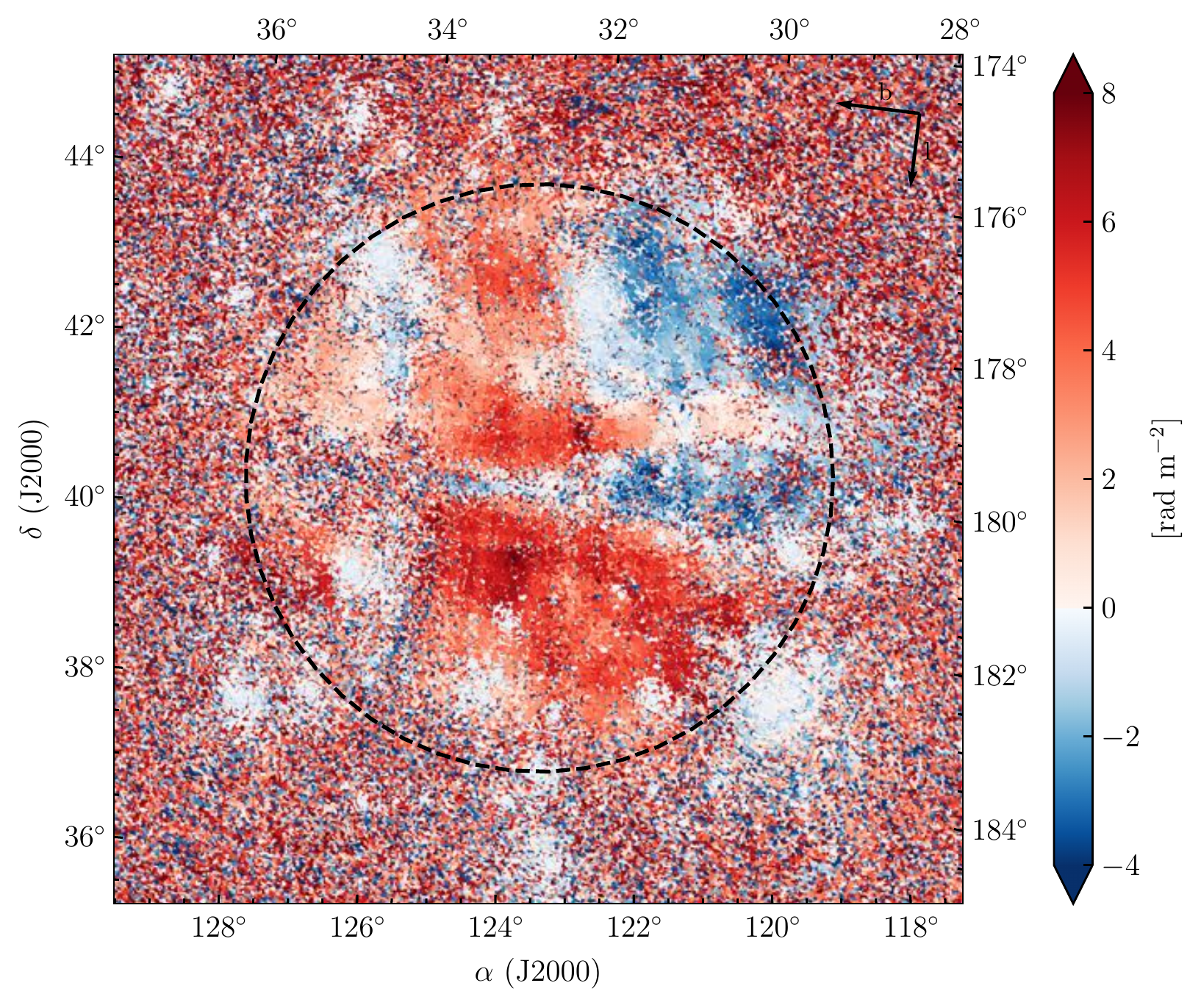}
\centering \includegraphics[width=.45\linewidth]{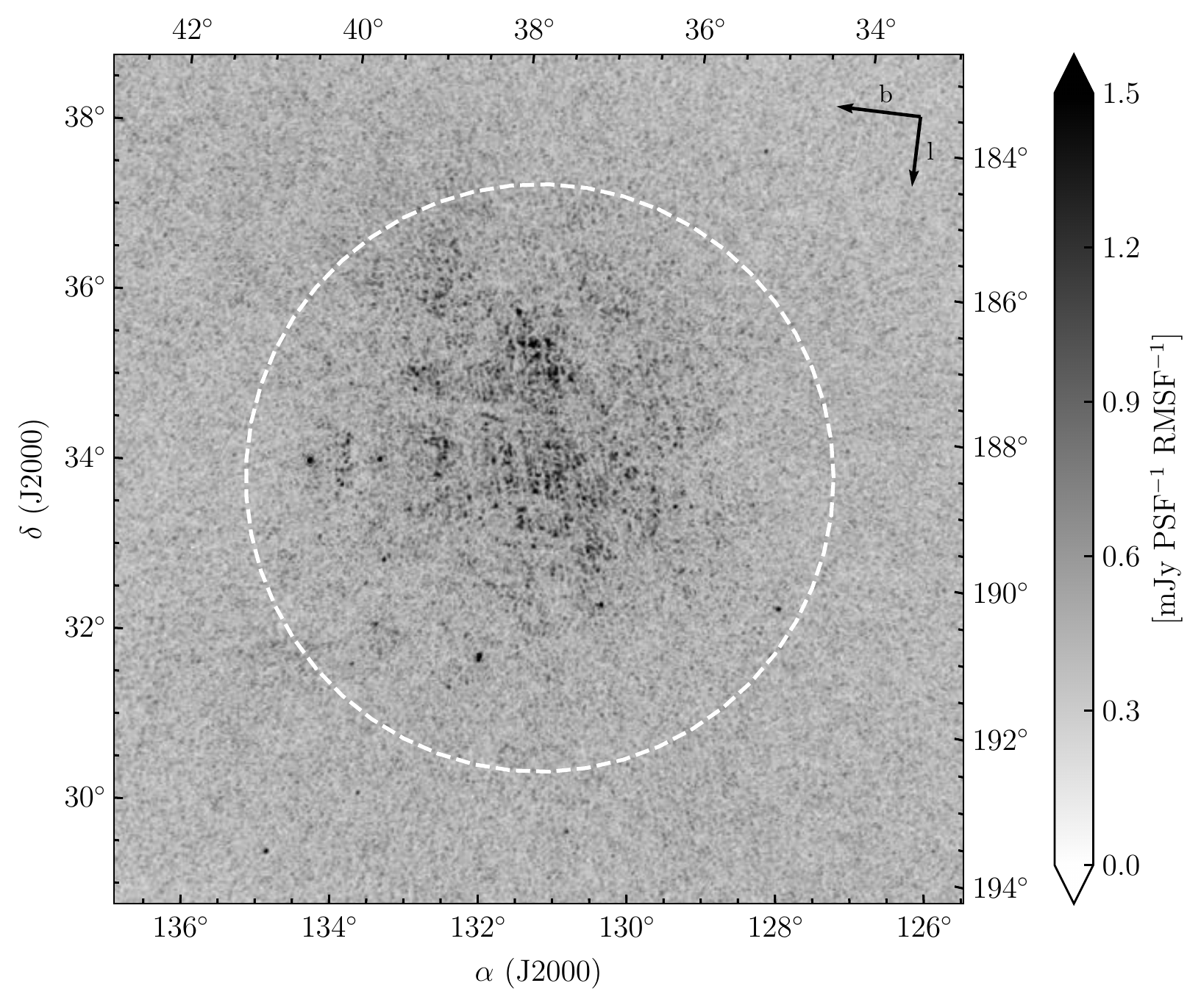}
\centering \includegraphics[width=.45\linewidth]{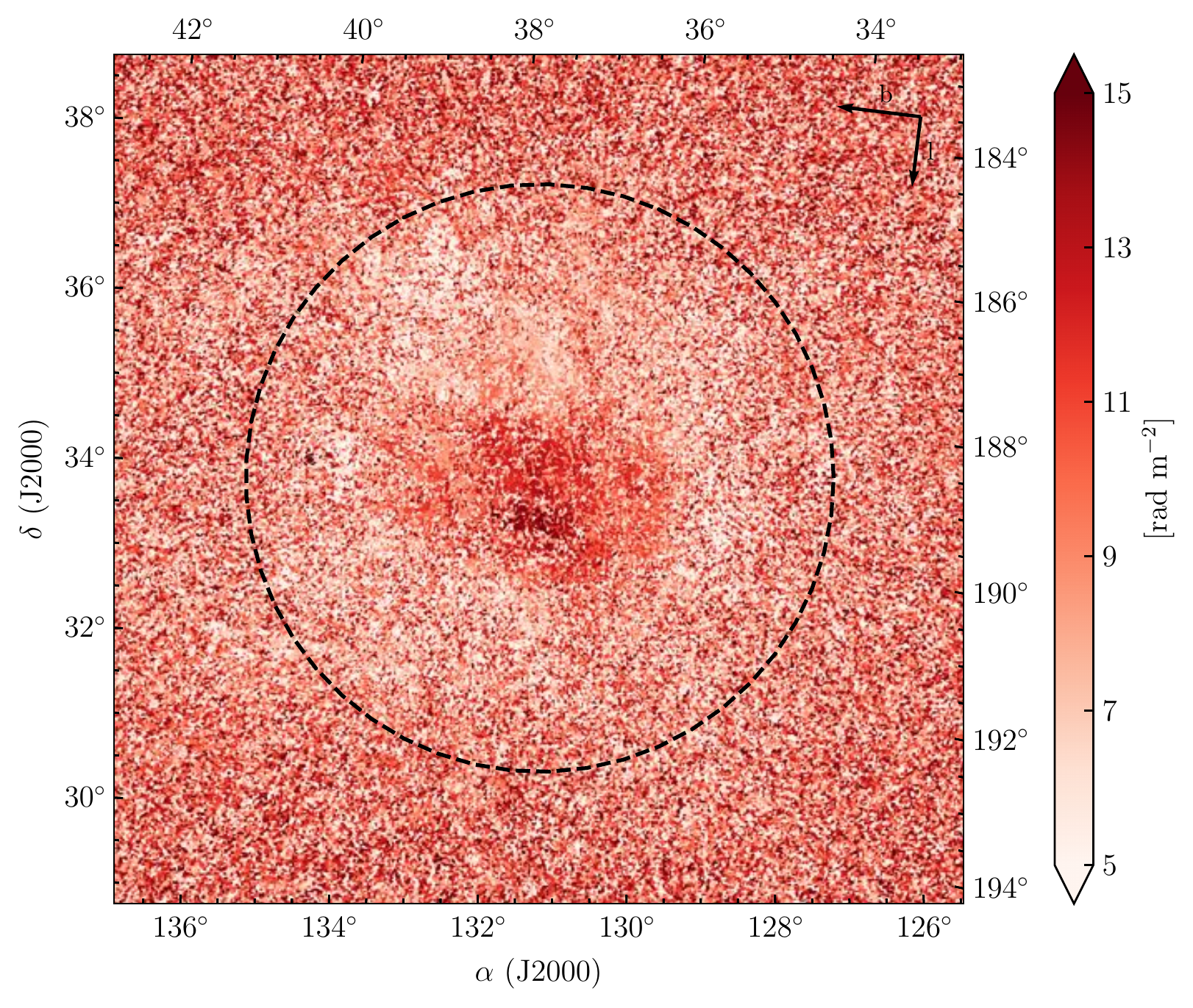}
\caption{Highest peak of the Faraday-depth spectrum in polarised intensity (left) and the map of the Faraday depth of the highest peak (right) for Fields A, B, and C from top to bottom, respectively. The dashed white and black circles have a diameter of $6.92^\circ$,  corresponding to the full width at tenth maximum of the primary beam at 150 MHz, which is the smallest primary beam in the analysed frequency range.}
\label{fig:Pmax_RMmap}
\end{figure*}

\section{Observational results}\label{Observational results}
\subsection{Faraday tomography of the diffuse polarised emission}
\label{Faraday tomography of diffuse polarised emission}
Diffuse polarised emission is detected in all three fields, but in each field, the emission shows different characteristics. It has a rich morphology in Fields A and B with the brightness of a few ${\rm mJy~PSF^{-1}~RMSF^{-1}}$. In contrast, the emission in Field C is very faint and patchy and has a diffuse morphology. The emission spans Faraday depths from $-6$ to $+8~{\rm rad~m^{-2}}$ in Field A, from $-4$ to $+8~{\rm rad~m^{-2}}$ in Field B, and from $+5$ to $+15~{\rm rad~m^{-2}}$ in Field C. To emphasise the most prominent features of the observed emission, we constructed images showing the highest peak of the Faraday-depth spectrum at each pixel (hereafter called the maximum polarised intensity image) and the Faraday depth of each peak. We present these images in Fig.~\ref{fig:Pmax_RMmap} for all three fields. Lack of emission at the edges of the fields is due to primary beam attenuation. 

\begin{figure*}[t]
\centering \includegraphics[width=.30\linewidth]{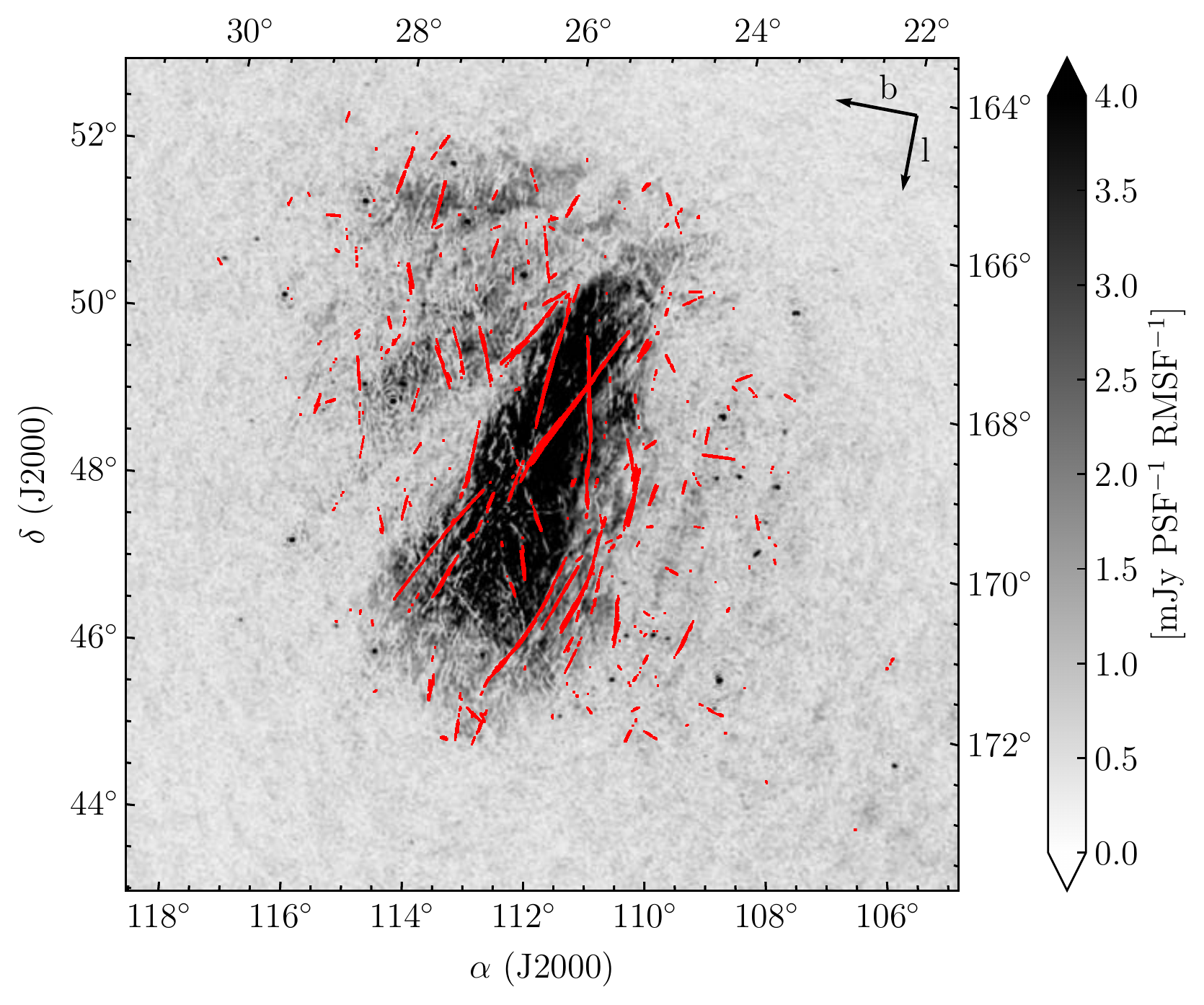}
\centering \includegraphics[width=.30\linewidth]{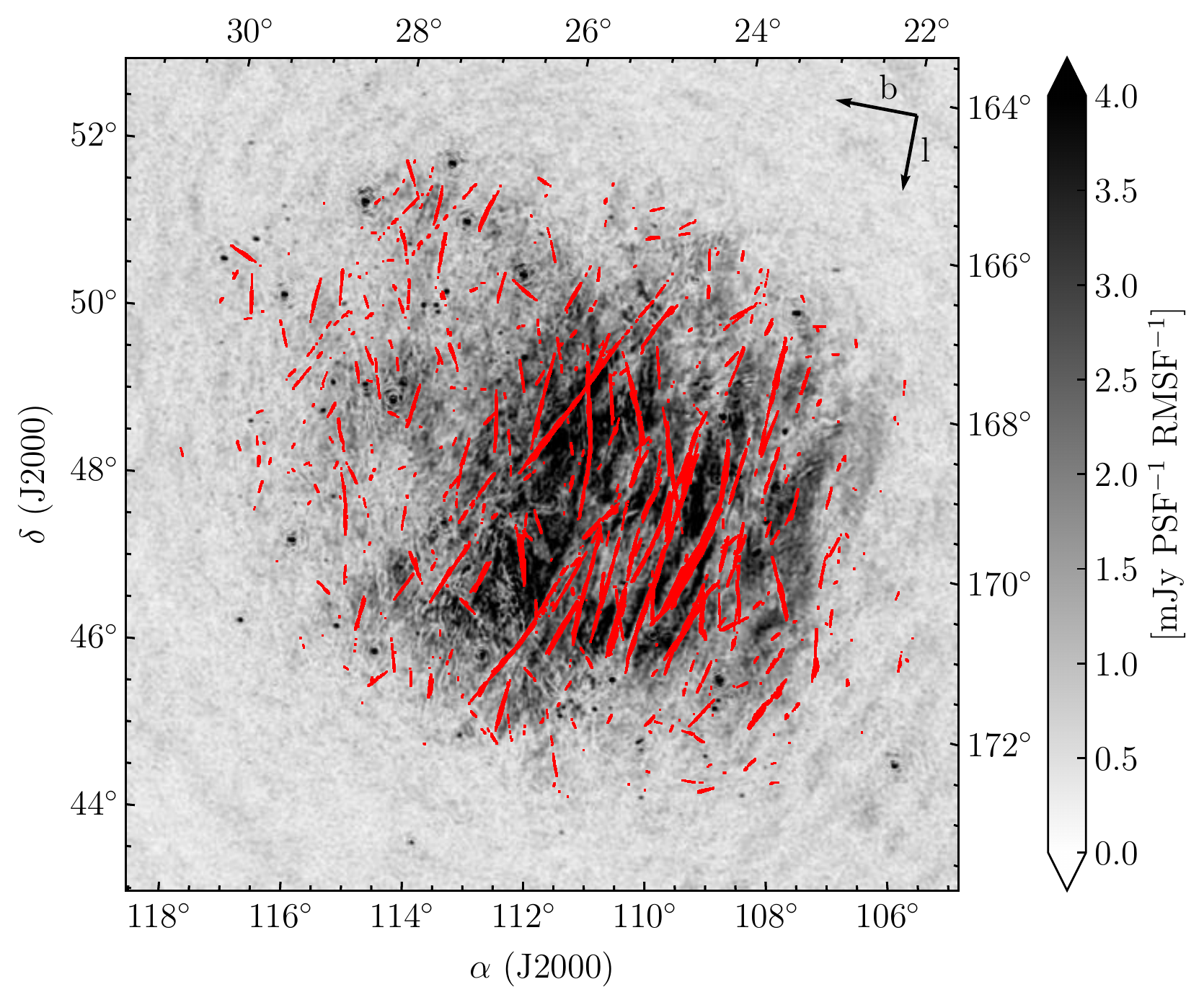}
\centering \includegraphics[width=.30\linewidth]{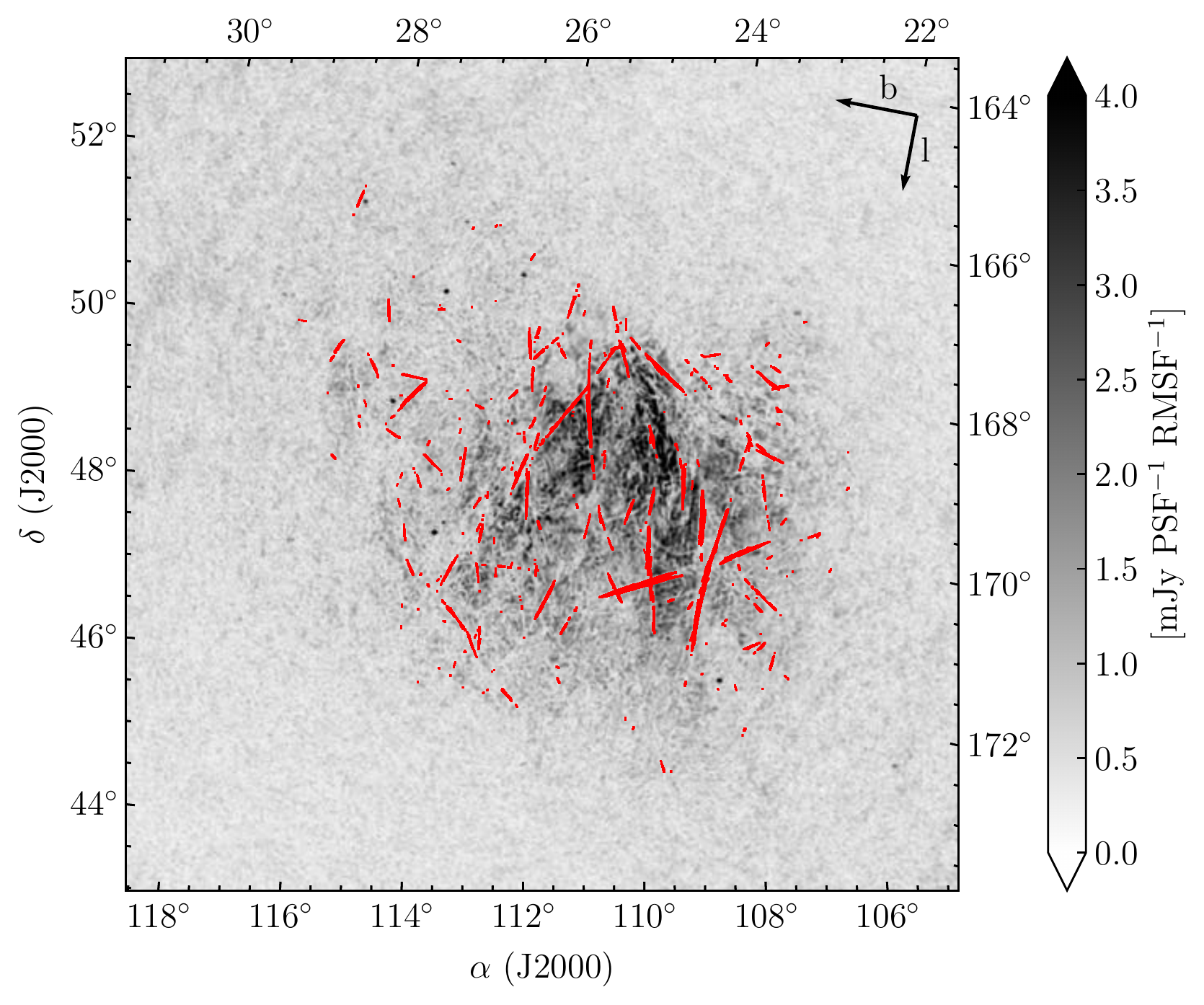}
\centering \includegraphics[width=.30\linewidth]{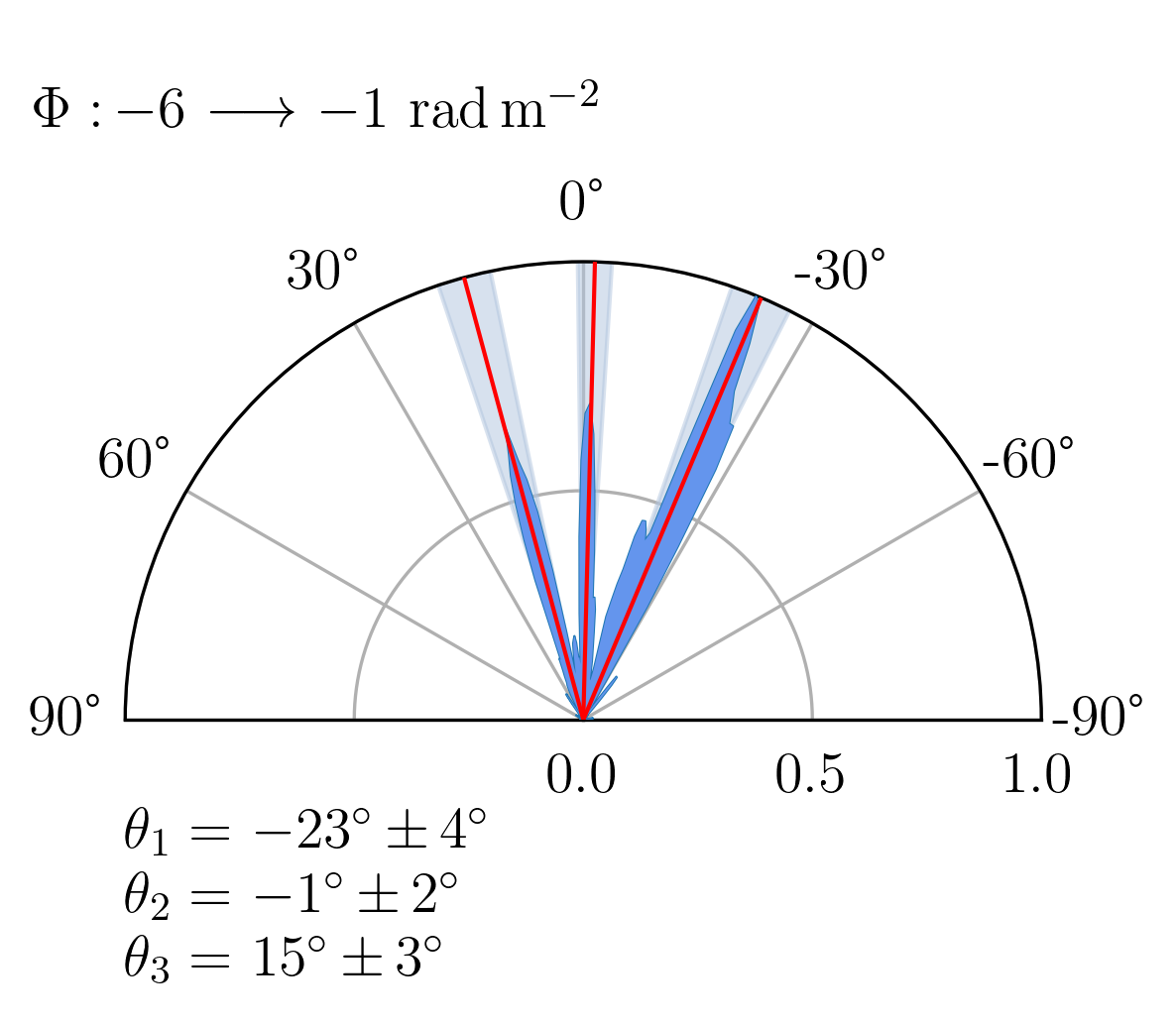}
\centering \includegraphics[width=.30\linewidth]{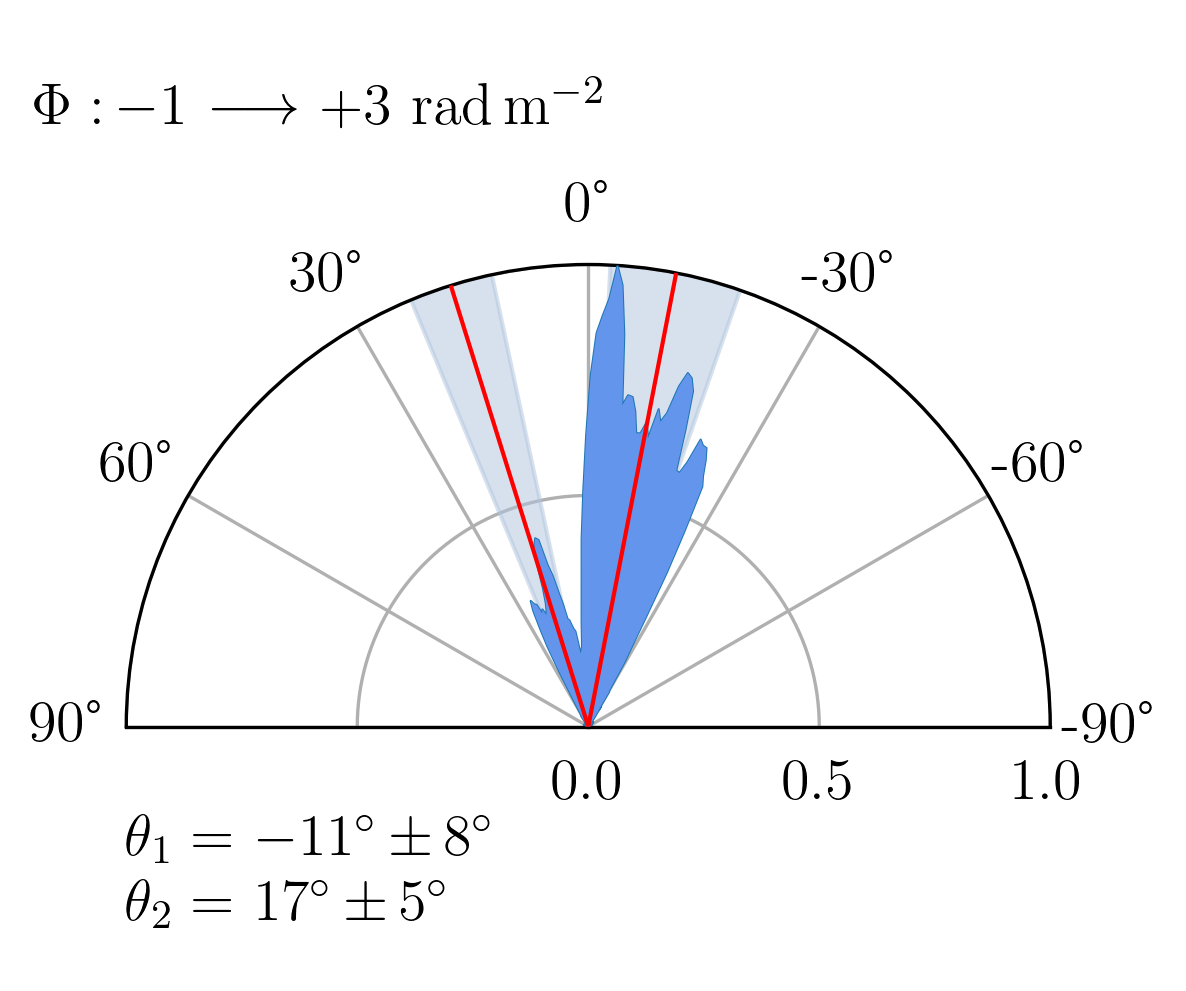}
\centering \includegraphics[width=.30\linewidth]{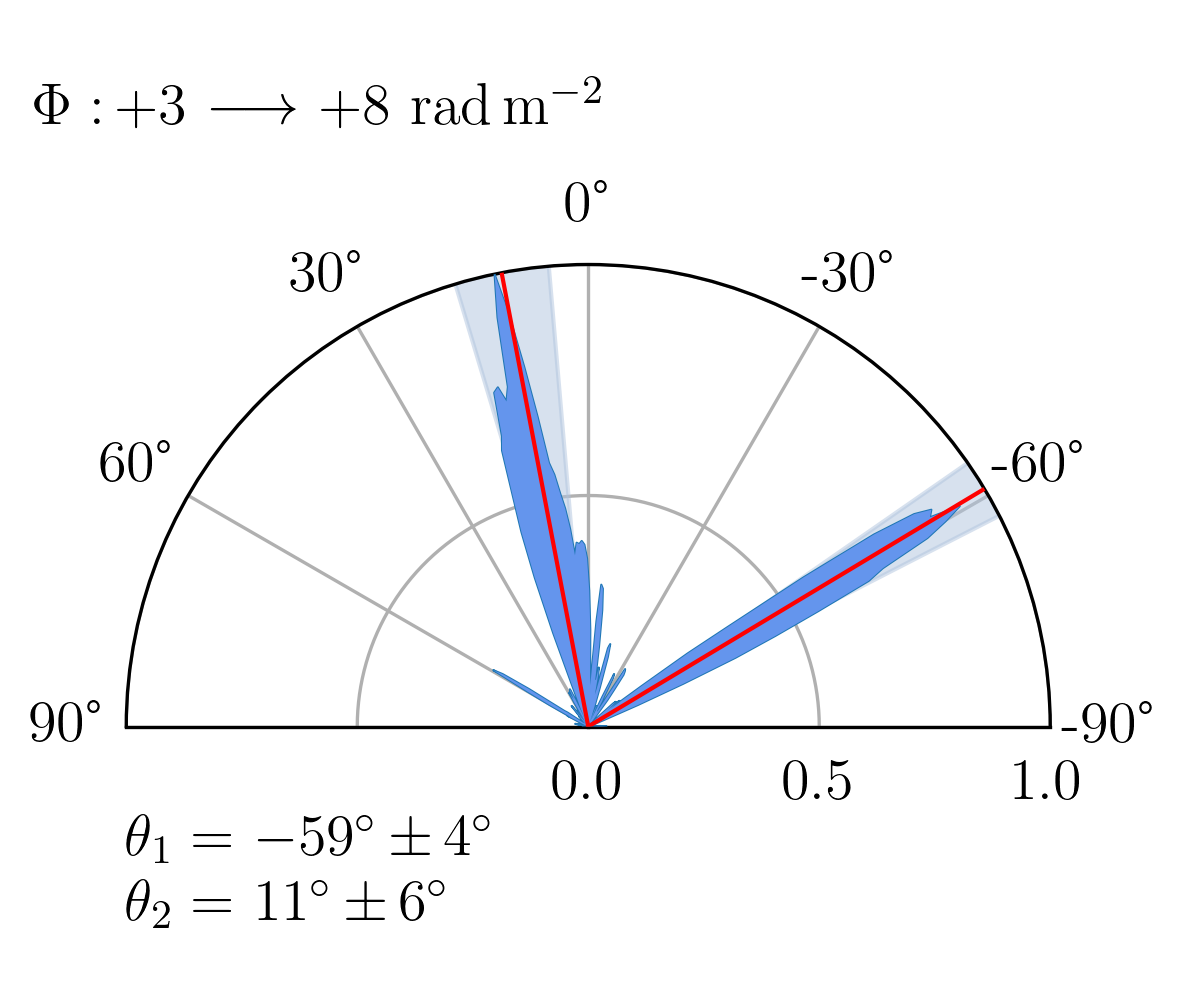}
\caption{RHT analysis performed on inverted maximum intensity images of Field A divided into three different Faraday-depth ranges. The upper part of the figure shows the results of the RHT analysis as weighted RHT back-projections (red lines) plotted over the maximum polarised intensity images. The lower part of the figure shows half-polar plots for the corresponding RHT back-projections, scaled according to their maximum. The Galactic plane orientation is at $0^{\circ}$. Red lines mark the averages of distinctive orientations, and light blue areas span the spread of the distributions for the corresponding averaged values.}
\label{fig:RHT_LOFAR_A}
\end{figure*}

The emission in Field A can be divided, based on its distinctive morphology, into three groups of structures that appear at different Faraday depths. This segmentation in Faraday depth is needed because these structures spatially overlap, as can be seen in the greyscale in Fig.~\ref{fig:RHT_LOFAR_A}. The first  morphological feature is an extended structure that appears in the centre of the image at a Faraday depth of $-6~{\rm rad~m^{-2}}$. As the Faraday depth increases towards $-1~{\rm rad~m^{-2}}$, the structure expands and first moves towards the north of the image and then mostly towards the southwest. The brightness temperature of this structure is about $4~{\rm mJy~PSF^{-1}~RMSF^{-1}}$. The second morphological feature is large-scale emission that covers almost the whole area of the image within the primary beam. This emission shows many coherent morphological structures and depolarisation canals, with a northwest-southeast orientation. The brightness of this emission is about $2.5~{\rm mJy~PSF^{-1}~RMSF^{-1}}$ and spans Faraday depths from $-0.75~{\rm rad~m^{-2}}$ to $+3~{\rm rad~m^{-2}}$. The third morphological feature is a boomerang-like structure in the southwest part of the image, which builds up from Faraday depths of $+3.25~{\rm rad~m^{-2}}$ and then disappears at $+8~{\rm rad~m^{-2}}$. The brightness of this structures reaches $2.5~{\rm mJy~PSF^{-1}~RMSF^{-1}}$.

The emission in Field B consists of distinctive patches with the brightness of about $1~{\rm mJy~PSF^{-1}~RMSF^{-1}}$. These patches first appear in the northwest part of the image at Faraday depth of $-4~{\rm rad~m^{-2}}$. As the Faraday depth increases towards $+8~{\rm rad~m^{-2}}$, the patches appear and disappear in different parts of the image. The patches at different Faraday depths seem to be organised in a large-scale cross-like structure (see images in the middle of Fig.~\ref{fig:Pmax_RMmap}).

The emission in Field C does not have any distinctive morphological features. The emission is very patchy and diffuse, with a brightness of $0.5~{\rm mJy~PSF^{-1}~RMSF^{-1}}$. It spans a Faraday-depth range from $+5~{\rm rad~m^{-2}}$ to $+15~{\rm rad~m^{-2}}$ (see Fig.~\ref{fig:Pmax_RMmap}).

Polarised emission in Fields A and B is observed at positive and negative Faraday depths. This implies that the magnetic field component parallel to the line of sight varies across each field in direction and probably in strength as well.  In Fig.~\ref{fig:Pmax_RMmap} positive Faraday depths (red regions) are associated with the magnetic field component parallel to the line of sight pointing towards the observer, while negative Faraday depths (blue regions) when it points in the opposite direction. This does not necessarily mean that the magnetic field has the same orientation along the line of sight. It only gives the dominant orientation of the magnetic field as probed by the polarised emission along the line of sight. In case the magnetic field has reversals, structures in Faraday spectra appear at both positive and negative Faraday depths. This is indeed the case for a large portion of Field A, where structures at positive and negative Faraday depths spatially overlap (see Fig.~\ref{fig:RHT_LOFAR_A}). This overlap of structures at positive and negative Faraday depths is seen only in a small fraction of Field B ($<5\%$). Polarised emission in Field C is observed only at positive Faraday depths, implying the magnetic field component pointing towards the observer across the whole field of view. 

The observed range of Faraday depths of polarised emission in the three fields is comparable to the range observed in the 3C 196 field \citep[from $-3$ to $+8~{\rm rad~m^{-2}}$; ][]{jelic15}. This suggests possible similar physical conditions contributing to the Faraday rotation in these fields. Furthermore, the large-scale magnetic field component is almost perpendicular to the line of sight in the 3C 196 field, making the plane-of-the-sky magnetic field component the dominant one \citep{jelic15}. This was later also supported in multi-tracer analyses of the field \citep{zaroubi15, jelic18}. We expect the same to be the case for the three fields studied in this paper, as they are within $20^\circ$ from the 3C 196 field and show polarised emission within a comparable Faraday-depth range as in the 3C 196 field.

\subsection{Brightness temperature and polarisation fraction}
We integrated the polarised intensity in the Faraday cubes along the Faraday depth axis to estimate the brightness temperature of the observed polarised emission in each field. The integrated brightness for each (RA, Dec) pixel is given by \cite{brentjens11},\begin{equation}
 PI=\frac{\Delta \Phi}{A_{\rm RMSF}}\sum^n_{i=1}{\left(PI(\Phi_i)-\sigma_{Q,U}\sqrt{\frac{\pi}{2}}\right)},    
\end{equation}
where $\Delta \Phi=|\Phi_{i+1}-\Phi_i|$ is an equidistant step in Faraday depth, $\sigma_{Q,U}\sqrt{\frac{\pi}{2}}$ is a correction for the non-zero mean of the noise in polarised intensity that is proportional to the noise in Stokes $Q$, $U$ RM cubes ($\sigma_{Q,U}$), and $A_{\rm RMSF}$ usually is the area of the restoring beam after deconvolution of the Faraday cubes with the corresponding RMSF \cite[e.g. RM-CLEAN,][]{heald09}. Our Faraday cubes are not deconvolved. Thus, instead of the restoring beam, we considered the area of the RMSF function from $-10$ to $+10~{\rm rad~m^{-2}}$. This is justified in our case as the peak brightness of the observed emission is only a few ${\rm mJy~PSF^{-1}~RMSF^{-1}}$ and the sidelobes of the RMSF function are of a few percent or smaller at $\Phi < -10~{\rm rad~m^{-2}}$ and $\Phi>+10~{\rm rad~m^{-2}}$. The sidelobe noise is therefore at these lower and higher Faraday depths comparable to or lower than the thermal noise.  Finally, for a PSF of $3.9\arcmin\times3.6\arcmin$, $1~{\rm mJy~PSF^{-1}}$ amounts to a brightness temperature of $1.3~{\rm K}$ at 130 MHz, a frequency that roughly corresponds to the weighted average of the observed $\lambda^2$ used in RM synthesis ($\lambda^2_0$).

\begin{figure}[!ht]
\centering
\centering \includegraphics[width=0.6\linewidth]{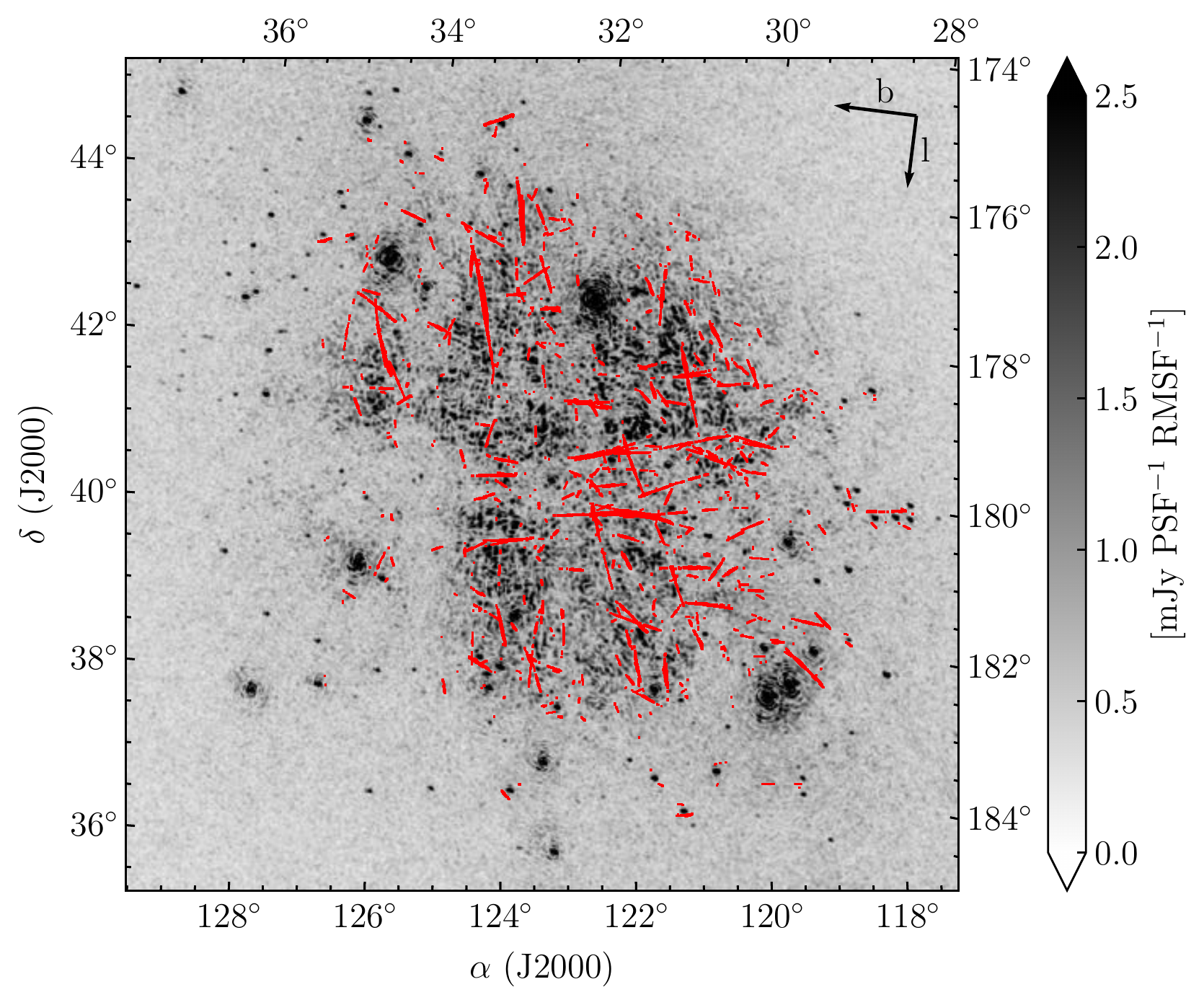}
\centering \includegraphics[width=0.6\linewidth]{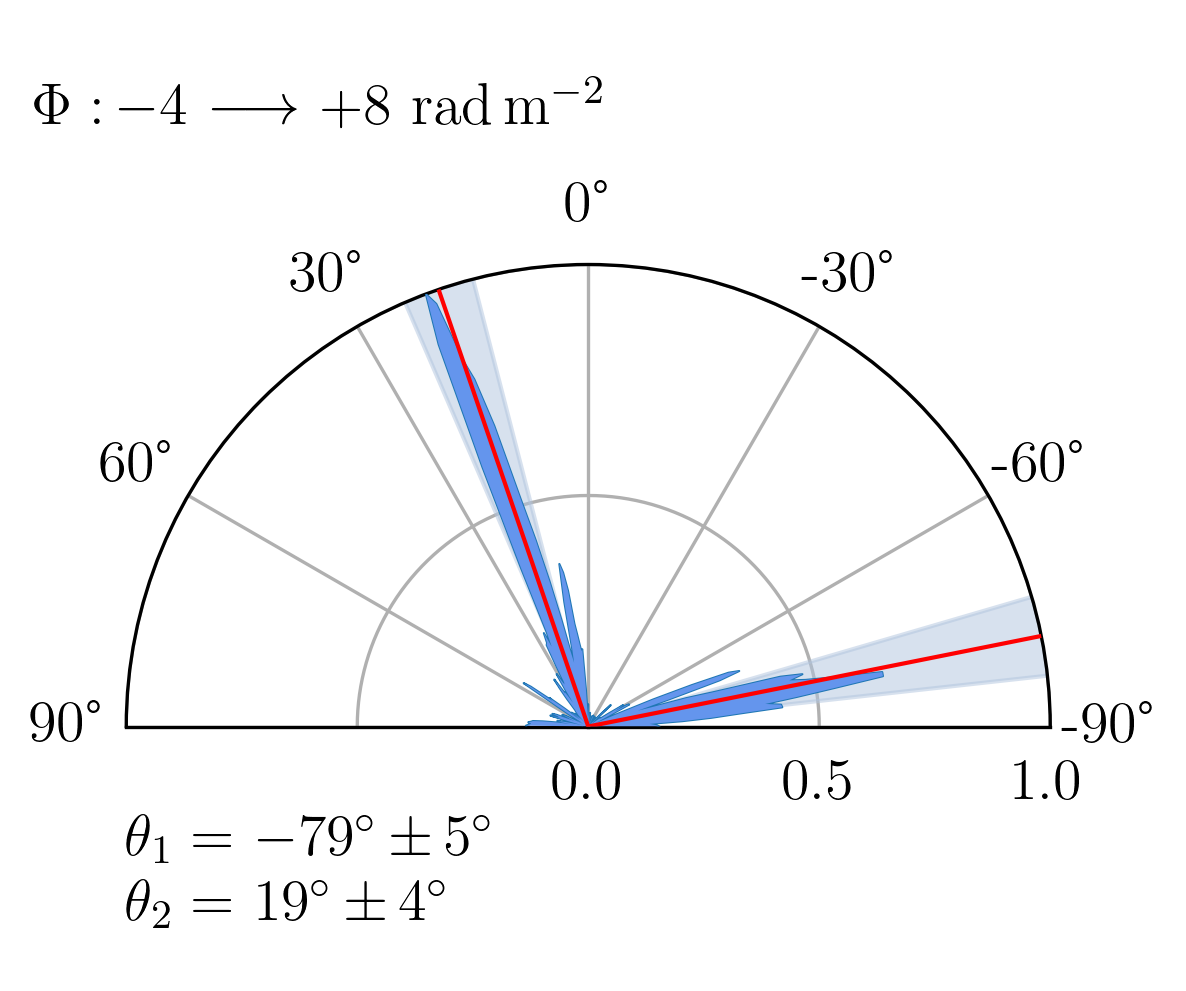}
\caption{Same as in Fig.~\ref{fig:RHT_LOFAR_A}, but for Field B.}
\label{fig:RHT_LOFAR_B}
\end{figure}

\begin{figure}[!ht]
\centering
\centering \includegraphics[width=0.6\linewidth]{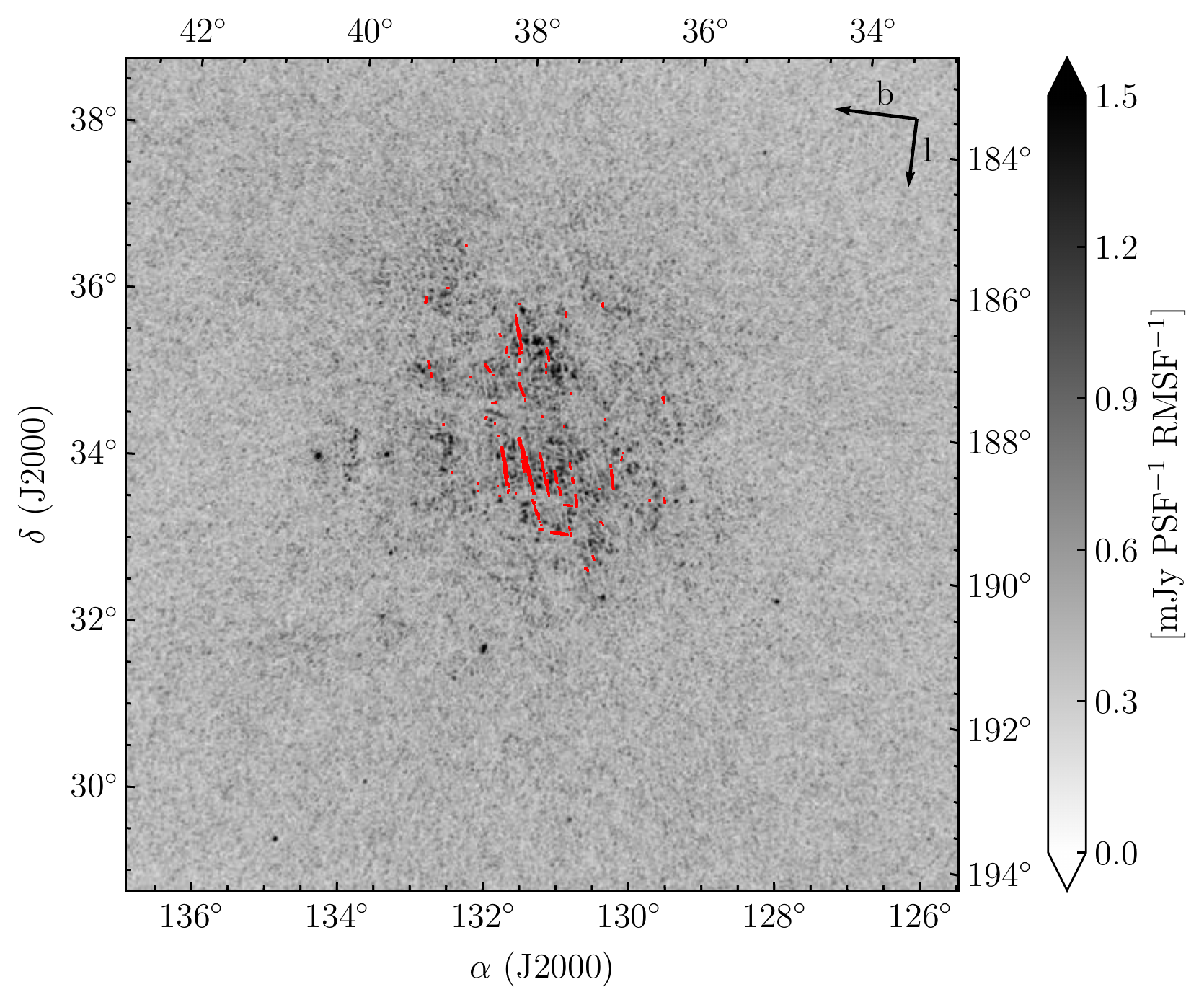}
\centering \includegraphics[width=0.6\linewidth]{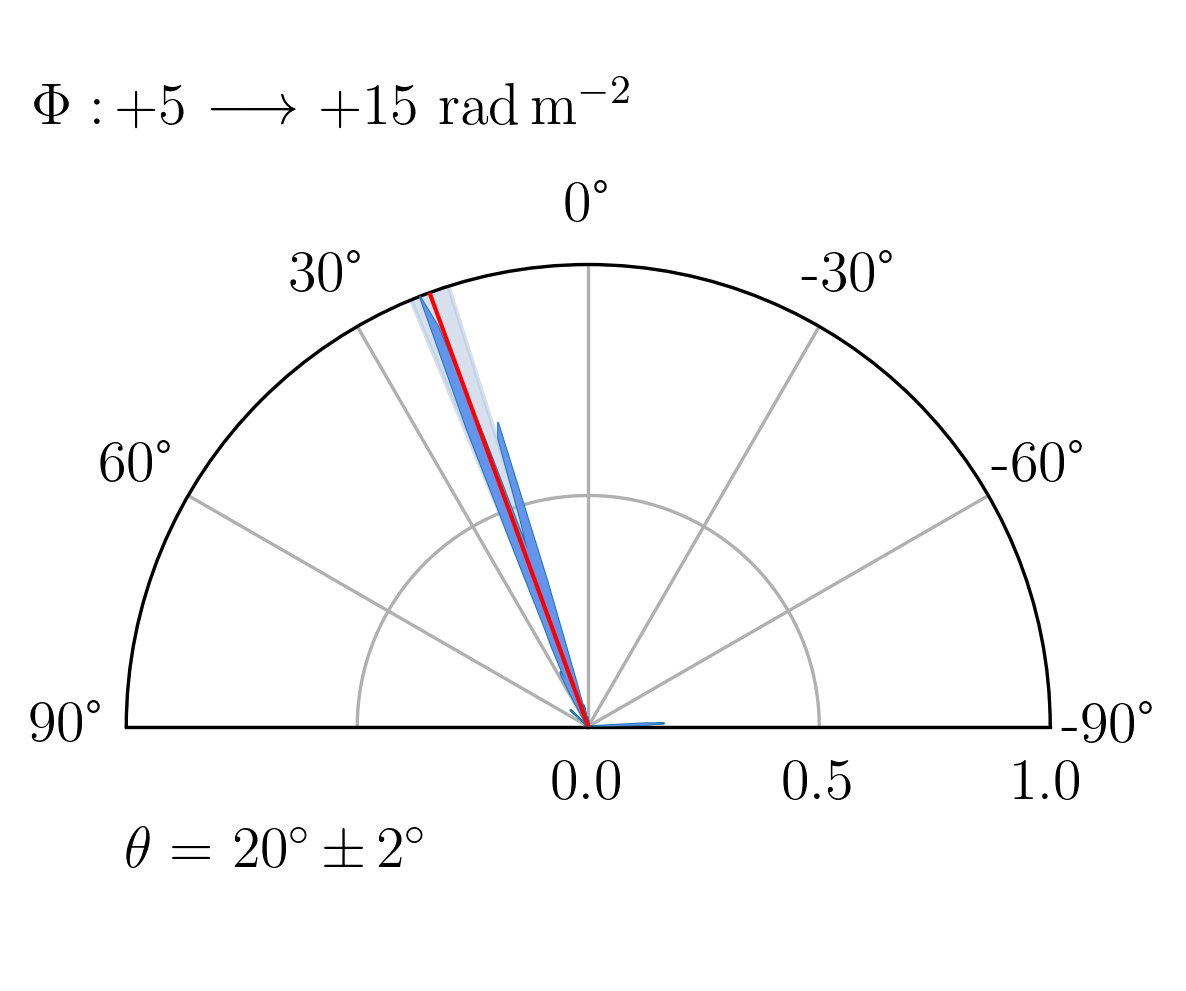}
\caption{Same as in Fig.~\ref{fig:RHT_LOFAR_A}, but for Field C.}
\label{fig:RHT_LOFAR_C}
\end{figure}

The average brightness temperature of the polarised emission in Field A is $5~{\rm K}$, in Field B, it is $1.5~{\rm K,}$ and in Field C, it is $0.6~{\rm K}$ (see Table~\ref{tab:polfrac}). To understand this difference, we estimated the total intensity and the polarisation fraction in each field. The polarisation fraction is defined as the ratio of the polarised and the total intensities. Because we have not observed emission in total intensity due to a lack of short baselines in our observations, we used the scaled Haslam 408 MHz All-Sky Map \citep{haslam81, haslam82, remazeilles2015}\footnote{\label{lcdm}\url{http://lambda.gsfc.nasa.gov}} as a proxy for the brightness temperature at $130~{\rm MHz}$. For the scaling, we use the spectral-index map of Galactic synchrotron emission between $45$ and $408~{\rm MHz}$ from \citet{guzman11}. The corresponding spectral indices, the brightness of the scaled emission at 408 MHz, and the calculated polarisation fractions are given in Table~\ref{tab:polfrac}. The polarisation fraction in Field A is $p=1.1 \% \pm 0.7 \%$, while Fields B and C show a lower polarisation fraction with values $p=0.5 \% \pm 0.3 \%$ and $p=0.2 \% \pm 0.1 \%$, respectively. These values should be taken as lower limits to the polarisation fraction at $130~{\rm MHz}$. LOFAR does not probe scales larger than a few degrees ($\gtrsim 3.5^\circ$), and there might be larger-scale polarised emission that we are missing. The lower polarisation fraction in Fields B and C arises because the depolarisation is stronger than in Field A. The same trend is also observed at 1.4 GHz (see also Table~\ref{tab:polfrac}), where the polarisation fractions are estimated based on the single-dish observations in total \citep[Stockert 25m + Villa Elisa 30m; ][]{Reich1981, Reich1986, Reich2001}\footnote{\label{mpifr}\url{https://www3.mpifr-bonn.mpg.de/survey.html}} and polarised intensity \citep[DRAO 26m, ][]{wolleben06}$^{\ref{mpifr}}$. The calculated polarisation fractions cannot be scaled and compared directly at the given frequencies because the data have different angular resolutions (36 arcmin at $1.4~{\rm GHz,}$ and 4 arcmin at $130~{\rm MHz}$). The higher-frequency data are more strongly affected by the beam depolarisation than the data at lower frequencies.

\begin{figure*}[!t]
\centering
\begin{subfigure}{0.26\textwidth}
\centering \includegraphics[width=\linewidth]{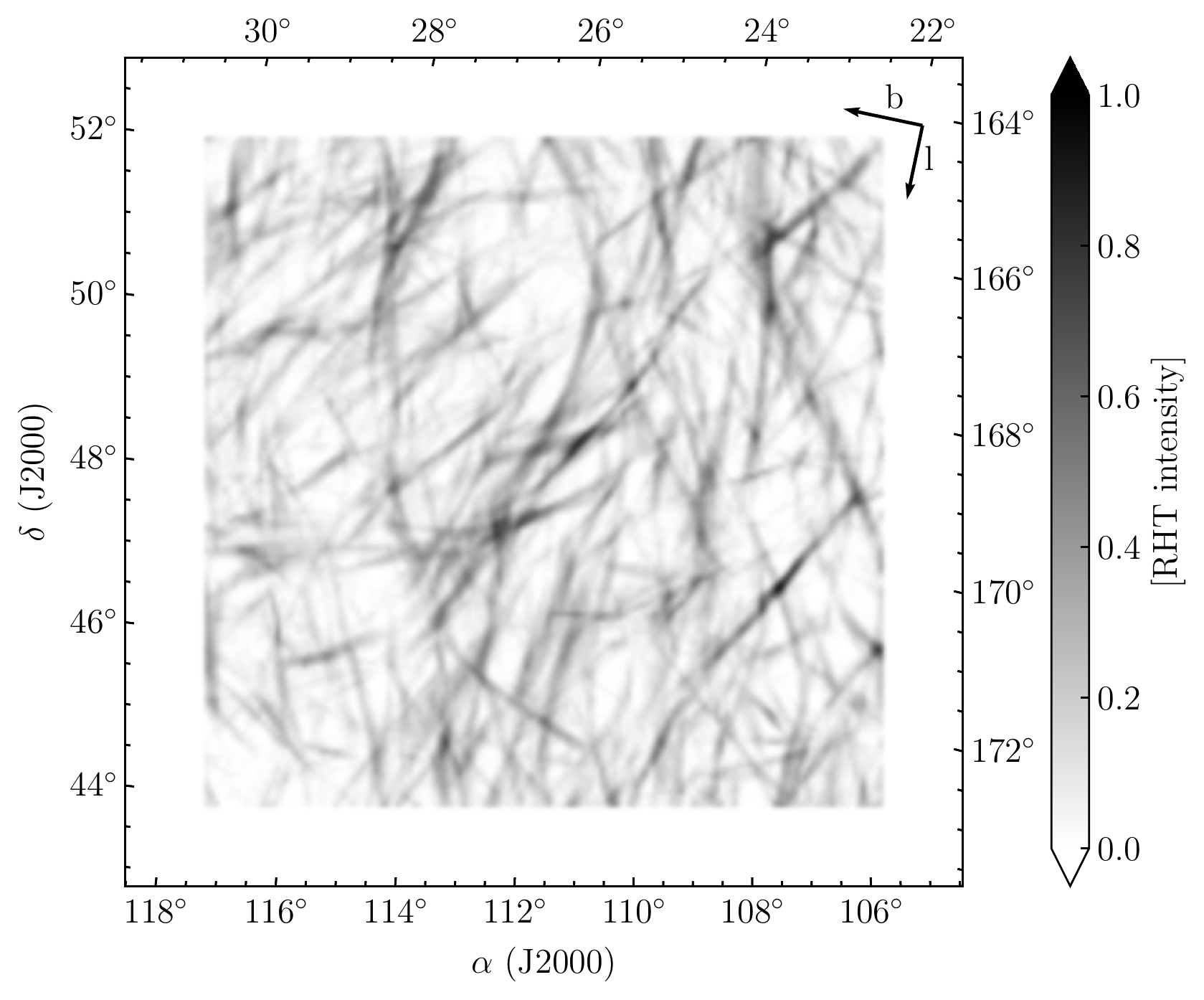}
\centering \includegraphics[width=\linewidth]{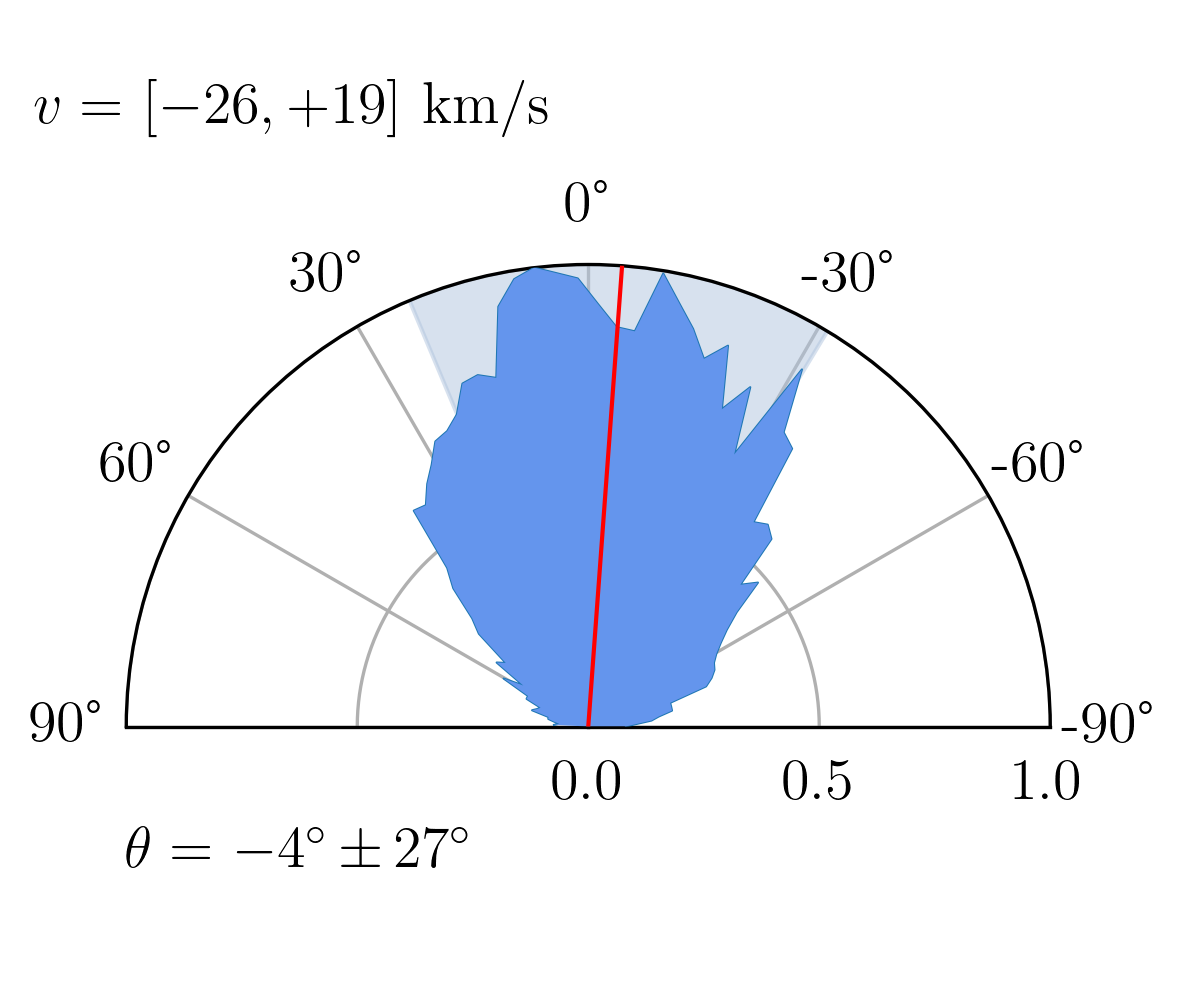}
\caption{Field A}
\label{fig:RHT_EBHIS_A}
\end{subfigure}
\begin{subfigure}{0.26\textwidth}
\centering \includegraphics[width=\linewidth]{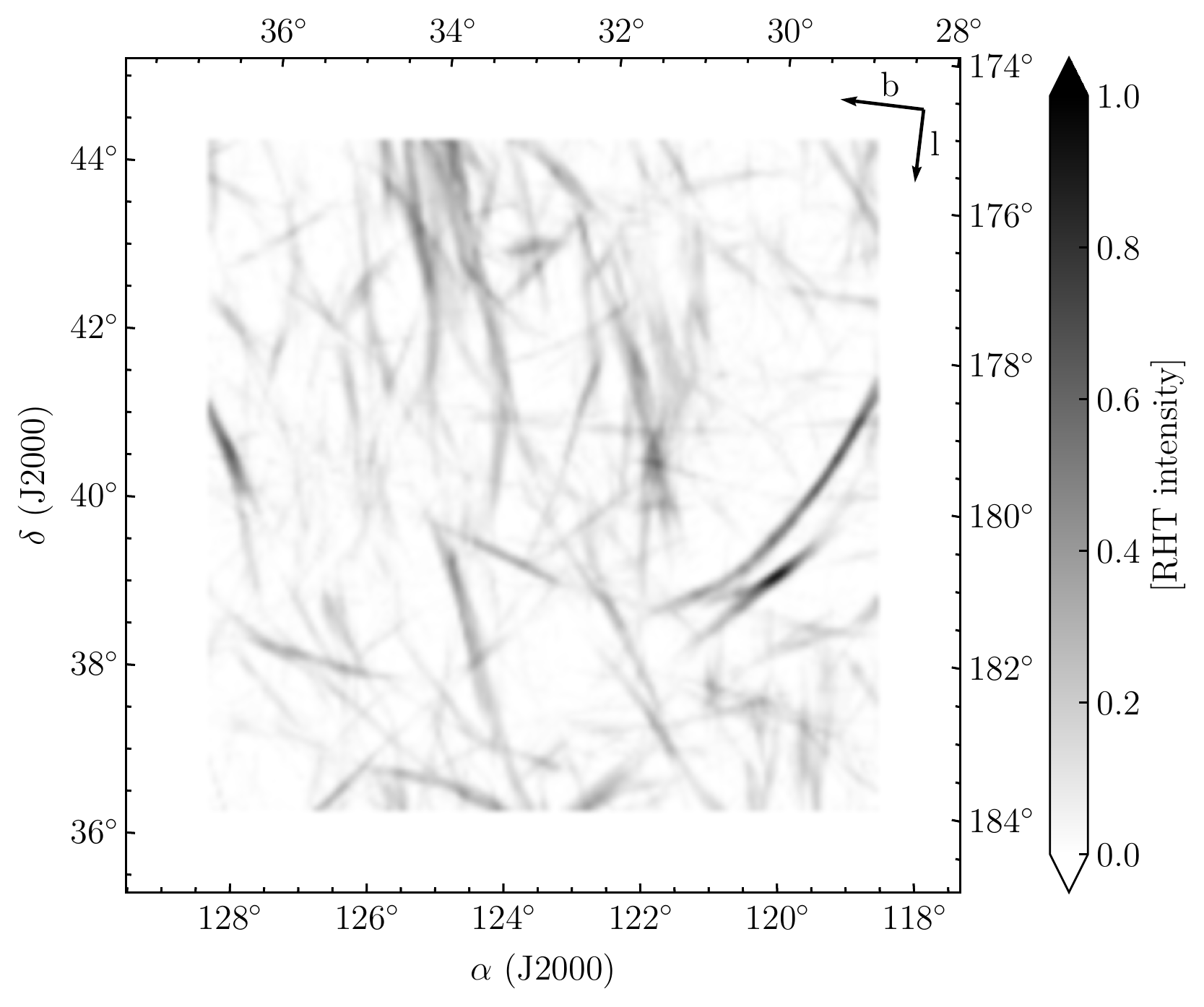}
\centering \includegraphics[width=\linewidth]{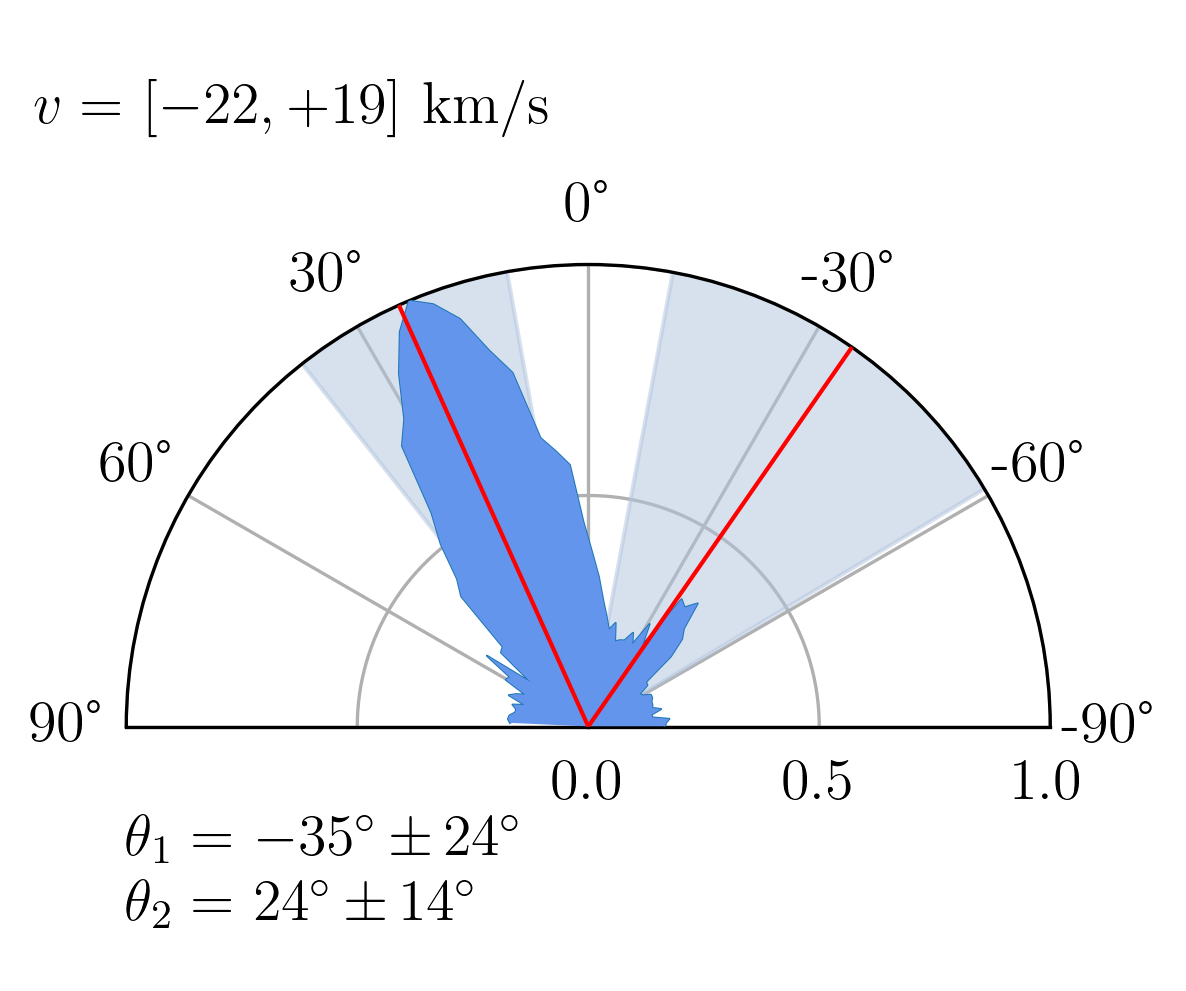}
\caption{Field B}
\label{fig:RHT_EBHIS_B}
\end{subfigure}
\begin{subfigure}{0.26\textwidth}
\centering \includegraphics[width=\linewidth]{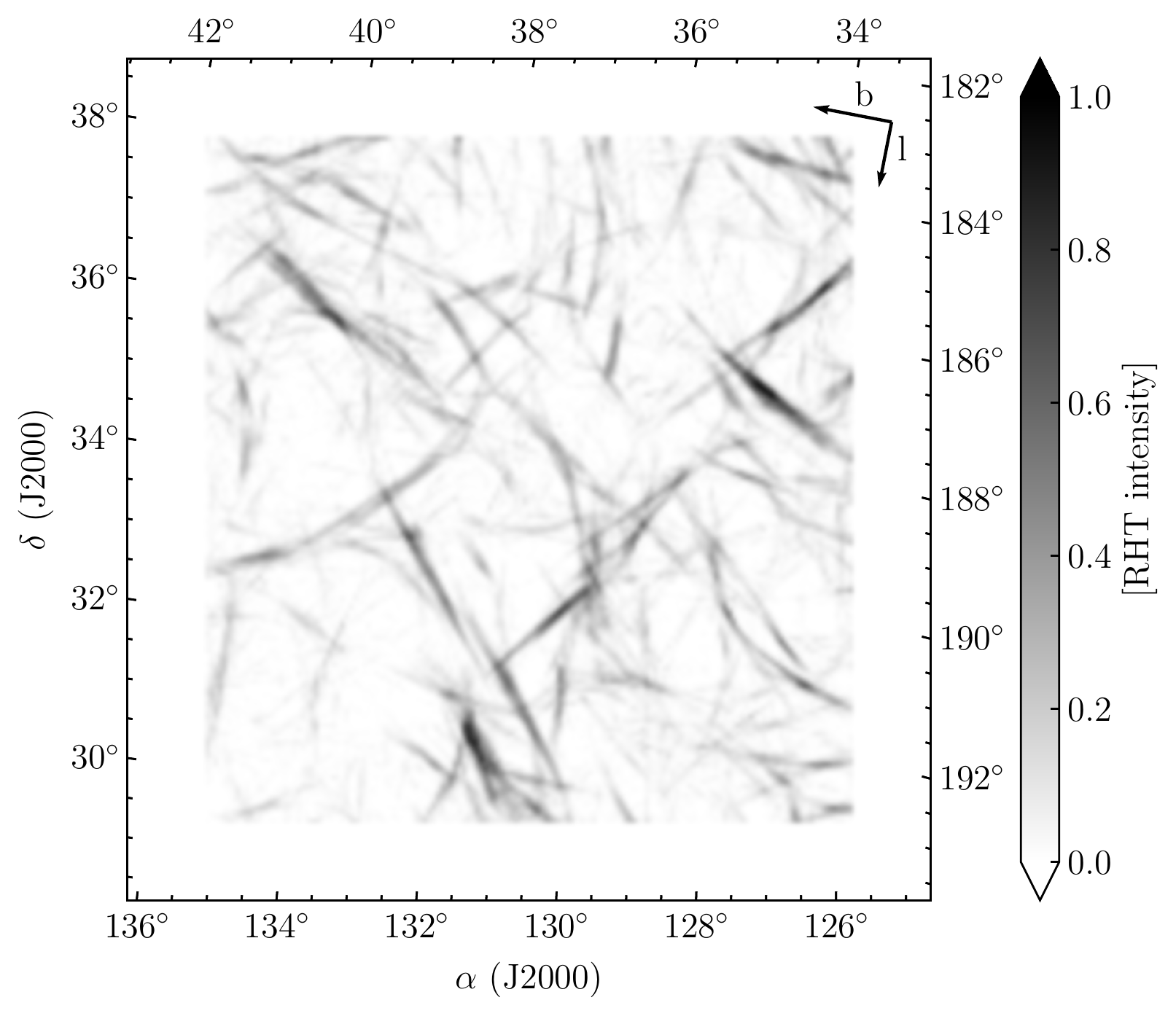}
\centering \includegraphics[width=\linewidth]{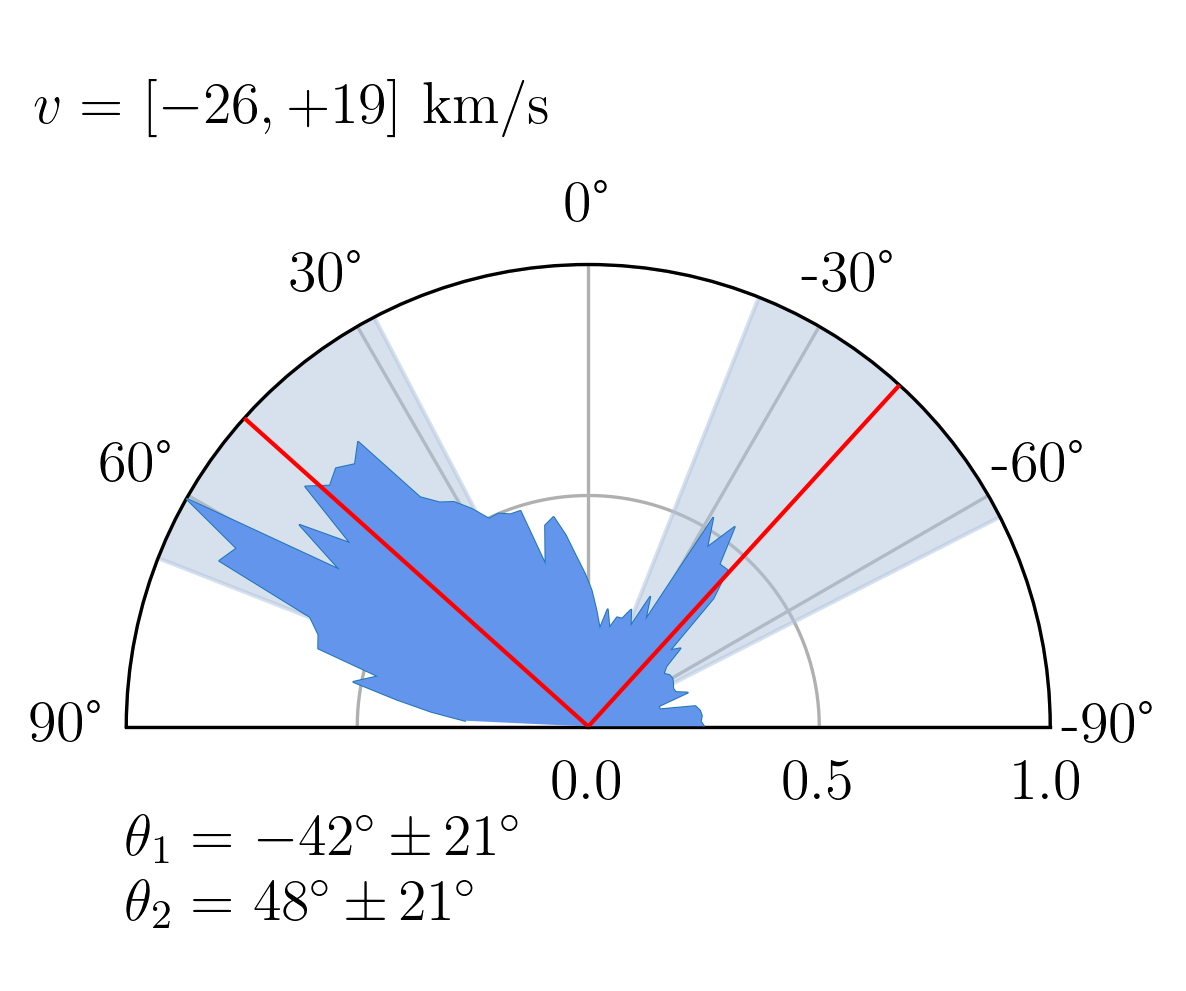}
\caption{Field C}
\label{fig:RHT_EBHIS_C}
\end{subfigure}
\caption{RHT analysis performed on the $\text{H}\textsc{I}$ brightness temperature at each velocity channel and integrated over the velocity range where the filaments are evident. The velocity range for each field is given below the images showing the RHT back-projections in the upper part of the figure. The lower part of the figure shows the corresponding distribution of the relative orientation of the $\text{H}\textsc{I}$ filaments with respect to the Galactic plane as in Figs.~\ref{fig:RHT_LOFAR_A}--\ref{fig:RHT_LOFAR_C}. The RHT input parameters are $D_K =10\arcmin$, $D_W =100\arcmin$ , and $Z=0.8$ (see main text).}
\label{fig:RHT_EBHIS_ABC}
\end{figure*}

\begin{figure*}[!t]
\centering
\begin{subfigure}{0.26\textwidth}
\centering \includegraphics[width=\linewidth]{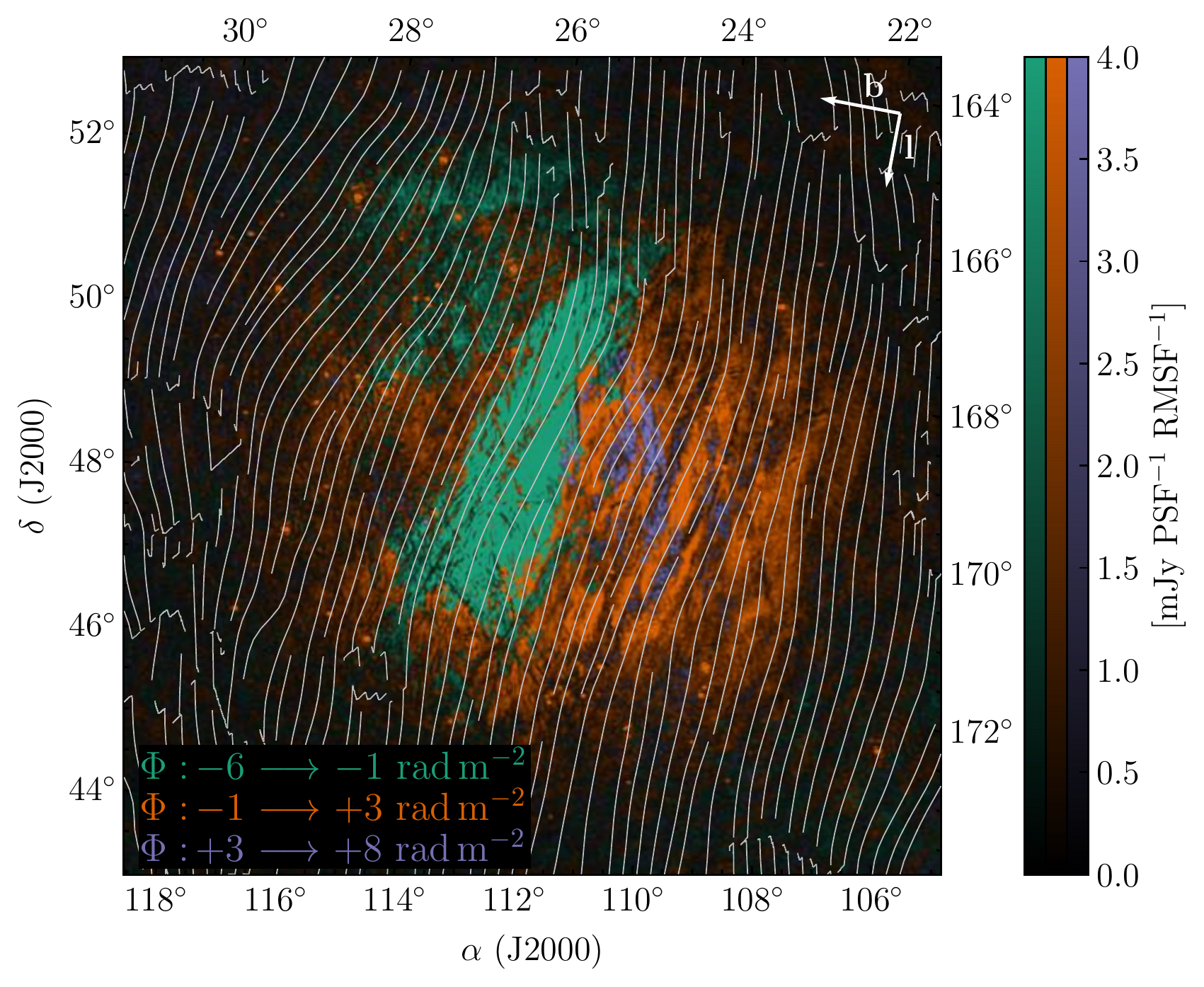}
\centering \includegraphics[width=\linewidth]{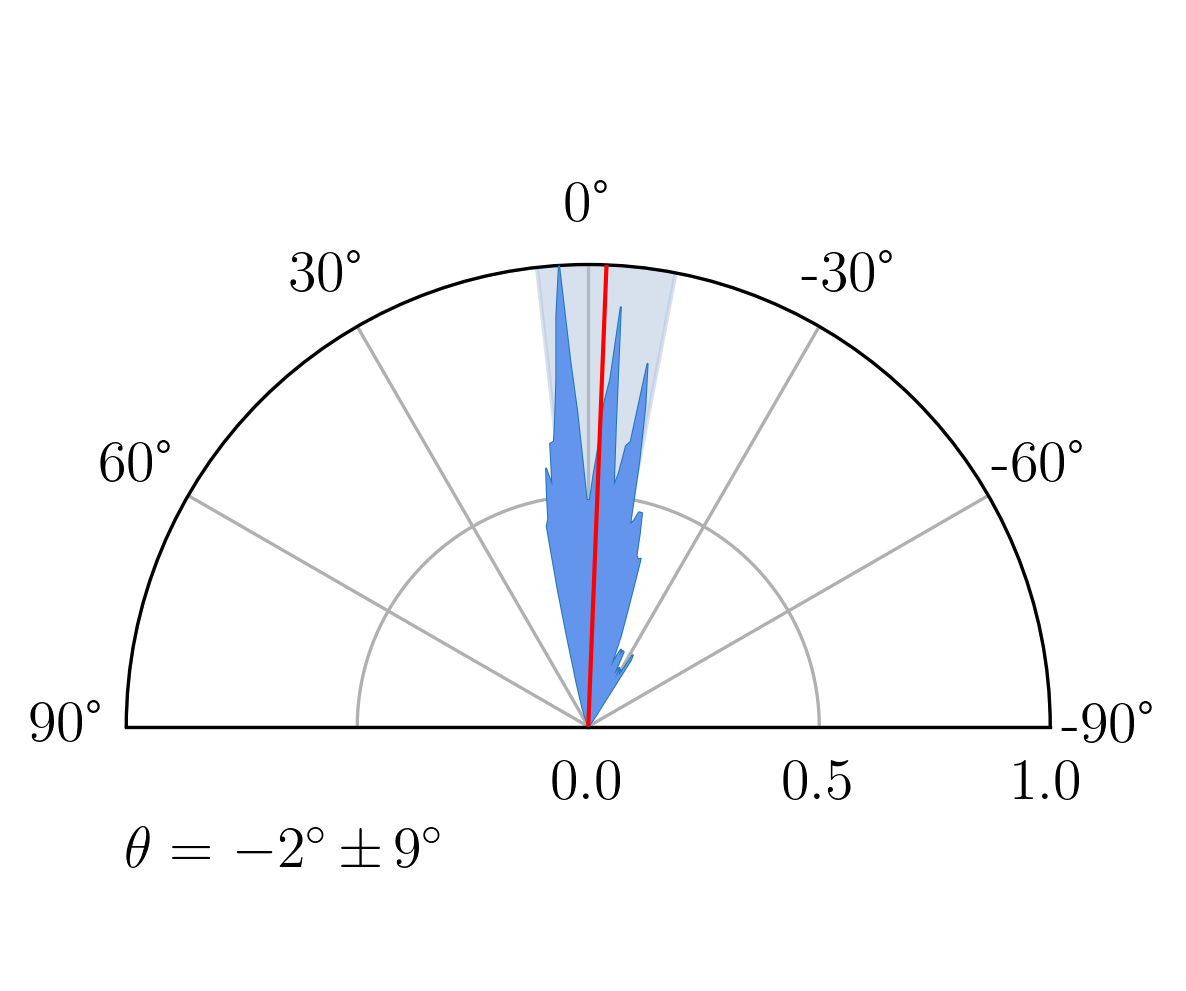}
\caption{Field A}
\label{fig:Planck_A}
\end{subfigure}
\begin{subfigure}{0.26\textwidth}
\centering \includegraphics[width=\linewidth]{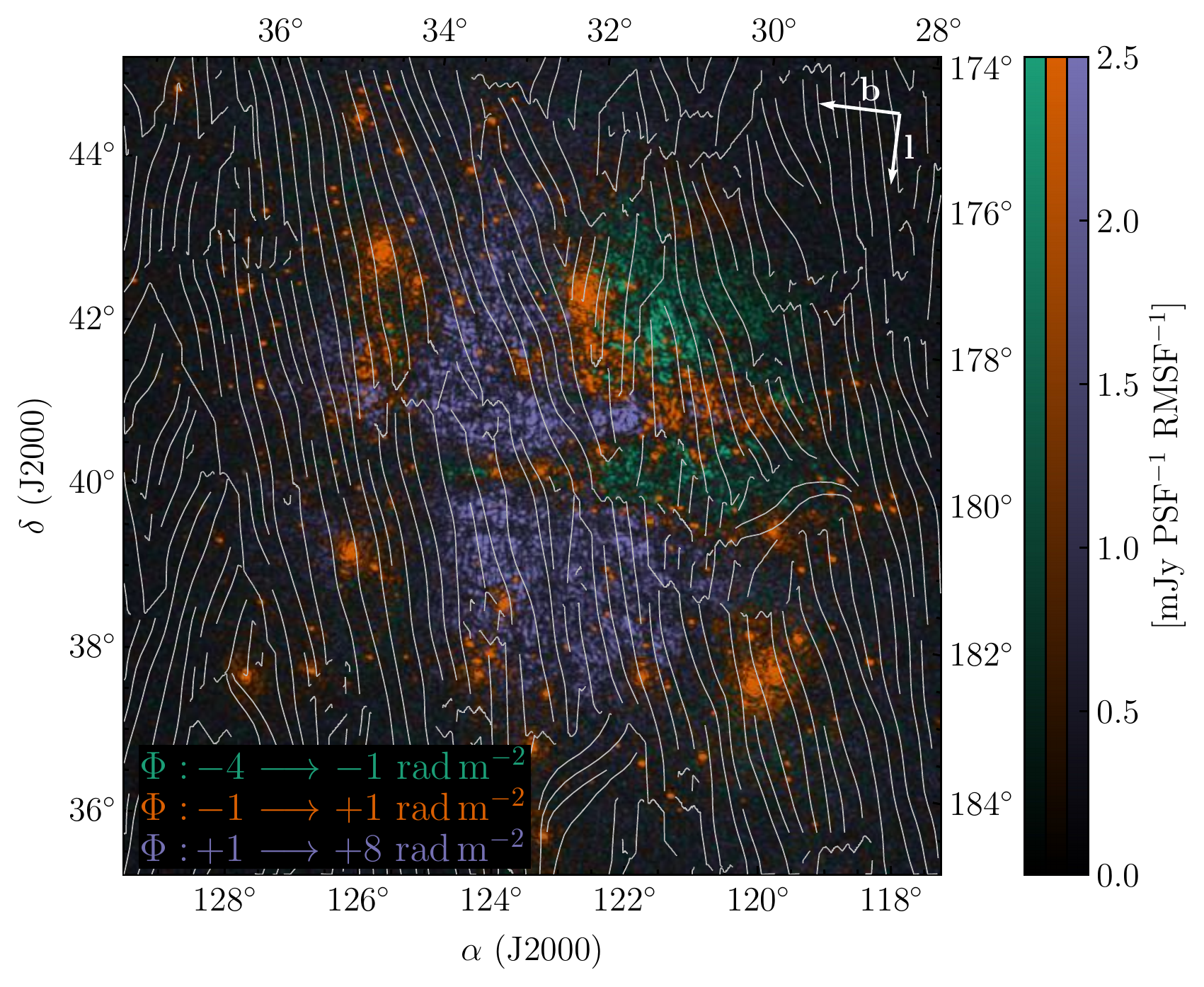}
\centering \includegraphics[width=\linewidth]{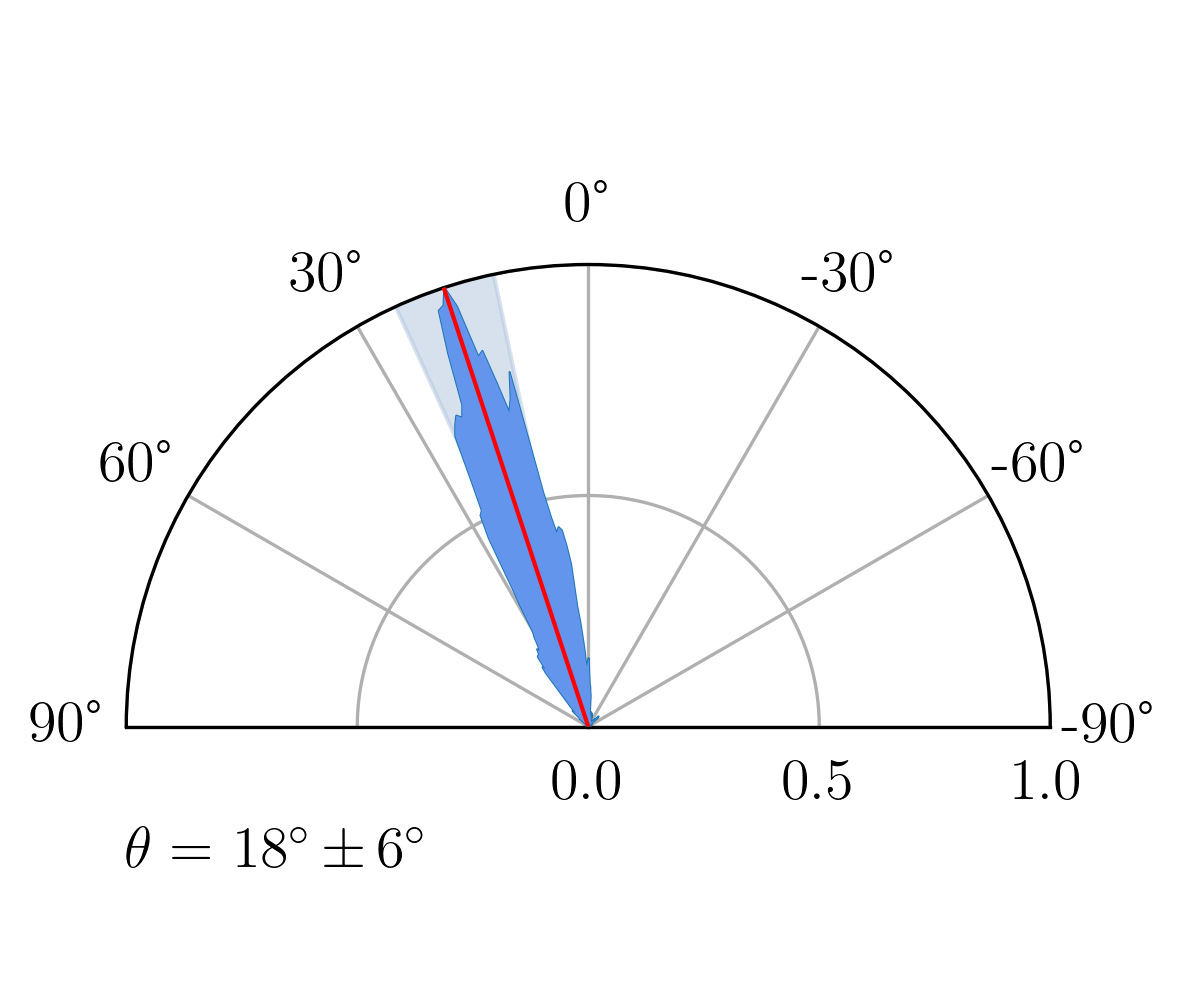}
\caption{Field B}
\label{fig:Planck_B}
\end{subfigure}
\begin{subfigure}{0.26\textwidth}
\centering \includegraphics[width=\linewidth]{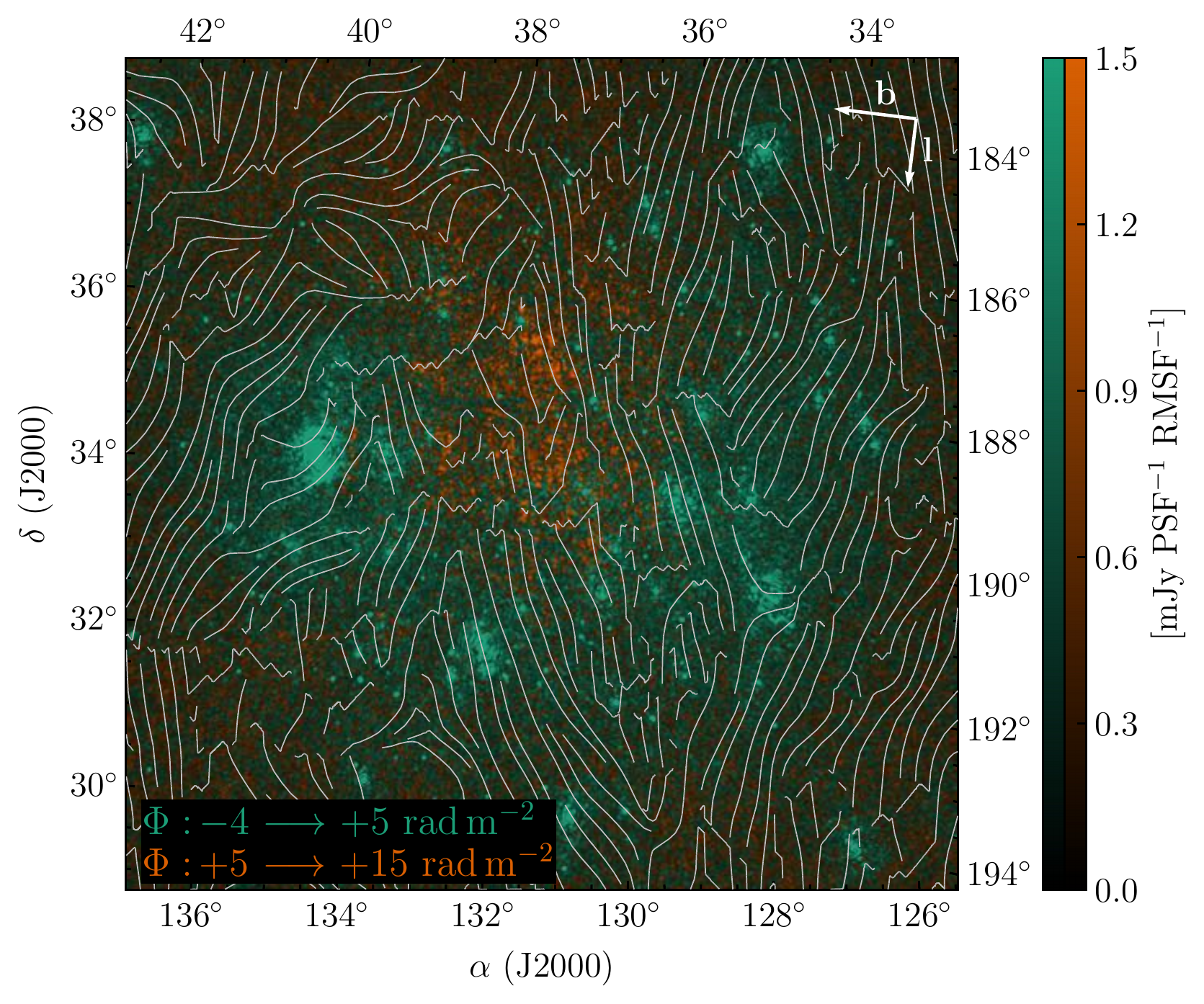}
\centering \includegraphics[width=\linewidth]{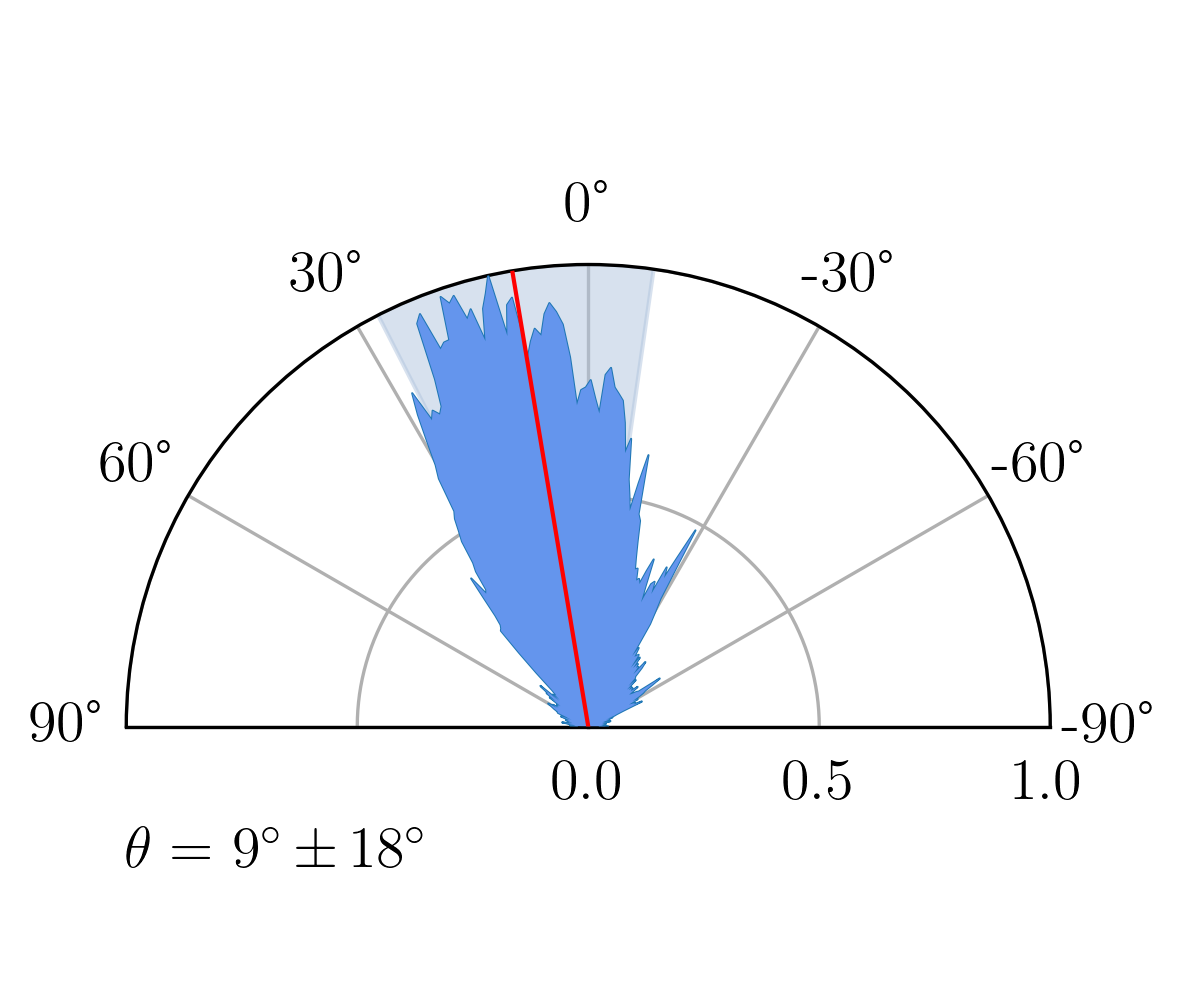}
\caption{Field C}
\label{fig:Planck_C}
\end{subfigure}
\caption{Top row: Rendering of the magnetic field lines on the plane of the sky from {\it Planck} over the LOFAR polarised emission sliced into different Faraday depth ranges as labelled in the figure. Bottom row: Corresponding distribution of the relative orientation of the plane-of-sky magnetic field lines with respect to the Galactic plane as in Fig.~\ref{fig:RHT_LOFAR_A}, for instance.}
\label{fig:Planck_ABC}
\end{figure*}

\section{Depolarisation canals}\label{sec:depolcanal}
Striking features that can be seen in the maximum polarised intensity images are the depolarisation canals (see the left side images in Fig. \ref{fig:Pmax_RMmap}). These are linear structures in the images showing no polarised emission, consistent with the noise. We mostly associate them with beam depolarisation at the location of discontinuities in the polarisation angle \citep{haverkorn04, jelic15}.  The most prominent canals are found in Fields A and B, and they are less abundant in Field C. We computed the orientation of these canals with respect to the Galactic-plane orientation as described in the following subsection.

\subsection{Orientation of the depolarisation canals}
To estimate the orientation of the depolarisation canals with respect to the Galactic plane, we followed \citet{jelic18} and used the rolling Hough transform \citep[RHT\footnote{\url{http://github.com/seclark/RHT}},][]{clark14}. The RHT is an algorithm for detecting straight lines in images. The output of the RHT is the quantity $R(\theta, x, y)$, where $\theta$ is the angle of the parametric straight line and $x$, $y$ are the pixel coordinates over the coherent linear structure. To visualise the result of the RHT, the back-projection $R(x, y)$ is obtained by integrating $R(\theta, x, y)$ over $\theta$. The RHT algorithm uses three input parameters. The first is the smoothing kernel diameter ($D_K$), which controls the suppression of the large-scale structures in the image; the second parameter defines the rolling window diameter ($D_W$); and the last parameter is the probability threshold ($Z$), which defines a minimum number of pixels (within the rolling window) in a particular direction $\theta$ needed for them to be considered as a part of the same linear structure \citep[for more details, see][]{clark14}. \citet{jelic18} showed that different combinations of the input parameters give very similar results for the orientation of the main linear depolarisation canals identified in the RHT back-projection, as the longest depolarisation canals dominate the overall distribution. The depolarisation canals in our fields have similar lengths as the depolarisation canals in the 3C 196 field. Therefore we used the same RHT input parameters ($D_K =8\arcmin$, $D_W =50\arcmin$ and $Z=0.8$) as in \citet{jelic18} and 
applied the RHT algorithm on inverted images of the maximum polarised intensity (1/image). This highlights the canals over the surrounding emission, as the linear structures that appear in emission are suppressed and the algorithm is only sensitive to the linear depolarisation canals that are enhanced in the inverted images. 

In addition to \citet{jelic18}, we introduced the weighting of the RHT back-projection to analyse only the canals that are surrounded by emission that is at least $7\sigma$ above the noise in polarised intensity. For each pixel in the maximum polarised intensity image, we calculated the mean value of emission in its surroundings within a region of $1\times3$ PSF. If this mean brightness was lower than the $7\sigma$ threshold, the identified linear structure connected to that pixel was not taken into account. This ensured that we did not analyse linear structures defined by very faint polarised emission, whose morphology is ambiguous and dominated by noise. Different threshold values used for the weighting give similar results for the mean orientation of the depolarisation canals, while the spread of the distribution around the mean is within $20\%$ compared to the results presented in the paper.

As mentioned in Sect.~\ref{Faraday tomography of diffuse polarised emission}, the emission in Field A can be divided into three different Faraday-depth ranges according to their morphologies. Therefore we performed the RHT algorithm separately for each range. The results of the RHT analysis for Field A are given in Fig.~\ref{fig:RHT_LOFAR_A}. The upper part of the figure shows the weighted RHT back-projections plotted over the maximum polarised intensity images for each Faraday-depth range. The lower part of the same figure presents the half-polar plots for the corresponding weighted RHT back-projections, quantifying the relative orientation of the depolarisation canals with respect to the Galactic plane. To calculate the mean and the spread of the distributions, we made a full-circle projection of the half-polar plot and visualised every point as a vector with length $R^2d\theta$. Integrating over the whole plane gives us a total vector whose direction and value measures the mean and the spread of the distribution \citep[for details, see][]{jelic18}. 
The first Faraday-depth range shows three prominent orientations of the depolarisation canals: $\theta_1 = -23^\circ \pm 4^\circ$, $\theta_2 = -1^\circ \pm 2^\circ$, and $\theta_3 = 15^\circ \pm 3^\circ$. The depolarisation canals have to a large extent negative orientations ($\theta_1$ and $\theta_2$). Two orientations ($\theta_2$ and $\theta_3$) are comparable in RHT intensities and have a lower intensity than the third orientation ($\theta_1$). Unlike the first, the second and third Faraday-depth ranges have two different orientations with respect to the Galactic plane, a negative  and a positive orientation. The second Faraday depth range has orientations with values of $\theta_1 = -11^\circ \pm 8^\circ$ and $\theta_2 = 17^\circ \pm 5^\circ$. In this case, the depolarisation canals have mostly negative orientations with a large spread. The third Faraday-depth range has orientations with values of $\theta_1 = -59^\circ \pm 4^\circ$ and $\theta_2 = 11^\circ \pm 6^\circ$. Here the two orientations are represented equally.

Finally, we also applied the RHT on Fields B and C. The results are given in Figs.~\ref{fig:RHT_LOFAR_B} and \ref{fig:RHT_LOFAR_C} and are presented as in Fig.~\ref{fig:RHT_LOFAR_A}. Two main orientations of the depolarisation canals with respect to the Galactic plane are detected in Field B, a negative and a positive orientation, with values of $\theta_1 = -79^\circ \pm 5^\circ$ and $\theta_2 = 19^\circ \pm 4^\circ$ , respectively. In the case of Field C, we find one positive orientation with a value of $\theta = 20^\circ \pm 2^\circ$.

\subsection{Comparison with the orientation of the HI filaments}
We compared the relative orientation of the depolarisation canals with the orientation of the $\text{H}\textsc{I}$ filaments using publicly available data from the Galactic Effelsberg–Bonn $\text{H}\textsc{I}$ Survey \citep[EBHIS\footnote{\url{http://cdsarc.u-strasbg.fr/viz-bin/qcat?J/A+A/585/A41}},][]{winkel16}. The EBHIS data have an angular resolution of $10.8\arcmin$ and a spectral resolution of $1.44~{\rm km~s^{-1}}$. We applied the RHT algorithm to the $\text{H}\textsc{I}$ brightness temperature at each velocity channel of the EBHIS data, with the RHT input parameters ($D_K =10\arcmin$, $D_W =100\arcmin$ and $Z=0.8$, following \citealt{clark14, jelic18}). We then summed the results over the velocity range in which the filaments are visible (see Fig~\ref{fig:RHT_EBHIS_ABC}). The results are robust to the variation in the RHT parameters and to the exact velocity range, as demonstrated and discussed by \citet{clark14}.

The upper part of Figure~\ref{fig:RHT_EBHIS_ABC} shows the RHT back-projections for the three fields, and the lower part presents the half-polar plots quantifying the relative orientation of the $\text{H}\textsc{I}$ filaments with respect to the Galactic plane. We took only the RHT results into account that can be found inside a circle corresponding to the LOFAR primary beam. 

The $\text{H}\textsc{I}$ filaments in Field A show a broad distribution of orientations ($\theta = -4^\circ \pm 27^\circ$), with a mean roughly parallel to the Galactic plane. The distribution encompasses the orientations of the depolarisation canals associated with the structures observed at Faraday depths of $[-6, -1]~{\rm rad~m^{-2}}$ and $[-1, +3]~{\rm rad~m^{-2}}$. Depolarisation canals at Faraday depths of $[+3, +8]~{\rm rad~m^{-2}}$ partially match the orientation of the $\text{H}\textsc{I}$ filaments. The RHT results for each velocity channel show that the $\text{H}\textsc{I}$ filaments have a dominant orientation of $-21^\circ\pm23^\circ$ over the velocities of $[-26, -15]~{\rm km~s^{-1}}$, then it becomes $12^\circ\pm19^\circ$ over the velocities of $[-14, -1]~{\rm km~s^{-1}}$, while at positive velocities of $[0, +13]~{\rm km~s^{-1}}$, their dominant orientation is $-24^\circ\pm25^\circ$. The dominant orientation of the depolarisation canals at Faraday depths of $[-6, -1]~{\rm rad~m^{-2}}$ and $[-1, +3]~{\rm rad~m^{-2}}$ seems to be consistent with the orientation of the $\text{H}\textsc{I}$ filaments at velocities of $[-26, -15]~{\rm km~s^{-1}}$ and $[0, +13]~{\rm km~s^{-1}}$. One of the two dominant orientations of depolarisation canals at Faraday depths of $[+3, +8]~{\rm rad~m^{-2}}$ ($11^\circ\pm6^\circ$) seems to be consistent with the orientation of the $\text{H}\textsc{I}$ filaments at velocities of $[-14, -1]~{\rm km~s^{-1}}$. However, it is very difficult to make definite conclusions given the broad distribution of the $\text{H}\textsc{I}$ filament orientations at different velocities that can overlap. In this field, we might be probing velocity-confused $\text{H}\textsc{I}$ gas, which makes the analysis more difficult. 

Unlike Field A, Fields B and C show two distinct orientations with respect to the Galactic plane. The $\text{H}\textsc{I}$ filaments in Field B have orientations of $\theta_1 = -35^\circ \pm 24^\circ$ and $\theta_2 = 24^\circ \pm 14^\circ$, while in Field C, they are  $\theta_1 = -42^\circ \pm 21^\circ$ and $\theta_2 = 48^\circ \pm 21^\circ$. The dominant orientation of the $\text{H}\textsc{I}$ filaments in Field B, $\theta_2$,  matches one of the two orientations of the depolarisation canals observed in the same field. The $\text{H}\textsc{I}$ filaments coherently show this orientation over a broad velocity range, between $-14~{\rm km~s^{-1}}$ and $+4~{\rm km~s^{-1}}$. At high negative velocities, between $-22~{\rm km~s^{-1}}$ and $-15~{\rm km~s^{-1}}$, the other orientation of the $\text{H}\textsc{I}$ filaments, $\theta_1$, dominates. There is no clear correspondence between $\theta_1$ orientation of $\text{H}\textsc{I}$ filaments and that of the depolarisation canals. In Field C, it is difficult to make a comparison because only a few depolarisation canals are detected.

On the one hand, the observed coherence of the $\text{H}\textsc{I}$ filament orientations over a wide range of velocities indicates a very uniform and ordered magnetic field. On the other hand, a change in their orientation at different velocities indicates a tangled line-of-sight magnetic field \citep{clark18}. To quantify this in our fields, we used publicly available velocity-integrated synthetic $\text{H}\textsc{I}$ Stokes parameter maps  ($I_{\text{H}\textsc{I}}$, $Q_{\text{H}\textsc{I}}$, $U_{\text{H}\textsc{I}}$)\footnote{\url{https://doi.org/10.7910/DVN/P41KDE}} derived by \citet{clark19} from the $\text{H}\textsc{I}$4PI all-sky spectroscopic data \citep{HI4PI16}, which combines the EBHIS data \citep{winkel16} in the north and the Parkes Galactic All-Sky Survey data \citep[GASS; ][]{McClureGriffiths2009} in the south. If the magnetic field is coherent and mostly in the plane of the sky, the synthetic polarisation fraction, defined as 
\begin{equation}
p_{\text{H}\textsc{I}} =\frac{ \sqrt{Q_{\text{H}\textsc{I}}^2+U_{\text{H}\textsc{I}}^2}}{I_{\text{H}\textsc{I}}},
\end{equation}
is expected to be high \citep{clark19}. After averaging $\text{H}\textsc{I}$-based Stokes parameters within the field of view of each field, we find values of $p=10.9 \%$ in Field A, $p=7.4 \%$ in Field B, and $p=5.6 \%$ in Field C. Values for Fields A and B are higher than the all-sky average of $p=6.7 \%$, while for Field C, the value is lower. We therefore conclude that the magnetic field is coherent and mostly in the plane of the sky in Fields A and B. 

\subsection{Comparison with the plane-of-the-sky magnetic field}\label{ssec:Planck}
We also compared the orientation of the depolarisation canals with the orientation of the plane-of-sky magnetic field component traced by the polarisation angle of dust-polarised emission rotated by $90^\circ$. The plane-of-sky magnetic field was obtained from the polarisation maps at $353~{\rm GHz}$ of the {\it Planck} satellite. The Stokes $Q$ and $U$ maps \citep{planckXIXinter, planck2015I} are publicly available at the Planck Legacy Archive\footnote{\url{http://pla.esac.esa.int}}. Due to the low signal-to-noise ratio of the observed polarised dust emission in all three fields, we used the {\it Planck} data smoothed to an angular resolution of $30\arcmin$ via the {\tt{HEALPix}}\footnote{\url{http://healpix.sourceforge.net}} package in python, \texttt{healpy} \citep{Gorski05,Zonca19}.

The results are shown in Fig.~\ref{fig:Planck_ABC}. The upper part of the figure shows visualisations of the magnetic field lines in the plane of the sky over the LOFAR polarised intensity as labelled by the different Faraday-depth ranges. The lower part of the figure presents the corresponding half-polar plots showing the relative orientation of the plane-of-the-sky magnetic field with respect to the Galactic plane. As in the case of the $\text{H}\textsc{I}$ filaments, we only took magnetic field lines within a central circle of the images into account that corresponded to the LOFAR primary beam.  

The mean orientation in Field A is slightly negative with a value of $\theta = -2^\circ \pm 9^\circ$, while in Fields B and C, it is positive with a value of $\theta = 18^\circ \pm 6^\circ$ and $\theta = 9^\circ \pm 18^\circ$, respectively. Given the calculated spread around the mean orientation, the plane-of-sky magnetic field is more ordered in Fields A and B than in Field C, as is clearly visible in the images in Fig.~\ref{fig:Planck_ABC}. Similar results were also obtained when we smoothed the data to $80\arcmin$, an angular resolution at which most of the intermediate Galactic latitude polarised dust emission has a signal-to-noise ratio higher than 3 \citep{Planck2018XII}.

In Fields A and B, most depolarisation canals have the same orientation as the plane-of-the-sky magnetic field component, which seems to be very ordered on the scale of the fields of view. Field B  also includes a distinct group of canals with orientations that depart from the plane-of-the-sky magnetic field by $\sim60^\circ$. This suggests the presence of two physical regions from which depolarisation canals may originate along the line of sight. Field C also shows a correspondence between the orientation of the plane-of-the-sky magnetic field component and the depolarisation canals, but the magnetic field is less ordered.

\begin{figure}[t]
\centering \includegraphics[width=.9\linewidth]{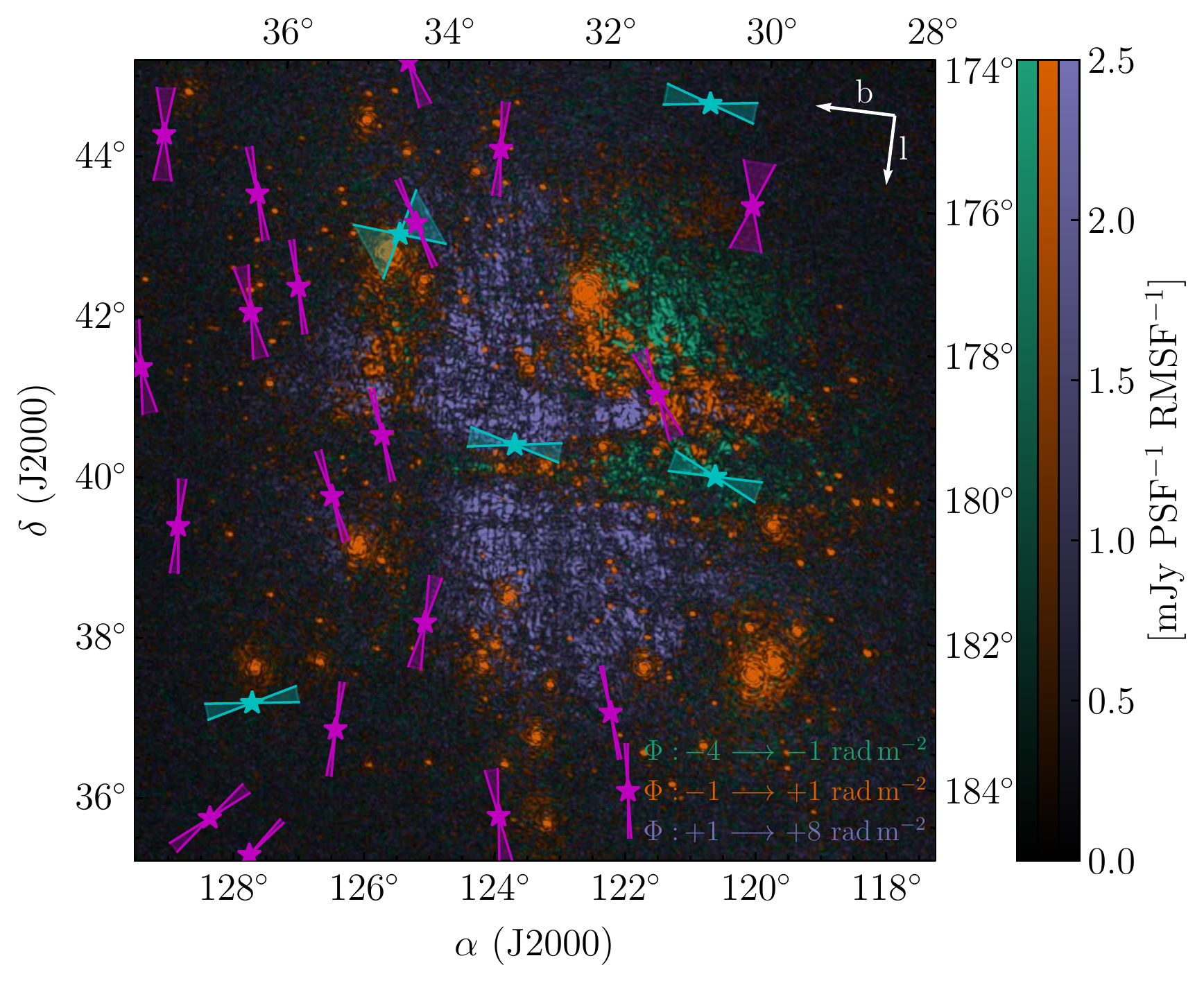}
\caption{Location of the stars in Field B. Colours correspond to two distinct groups of stars that trace different magnetic field orientations (cyan for stars at a distance closer than 200 pc, and magenta for stars at distances larger than 200 pc). The measured starlight polarisation angle is illustrated by two segments, whose aperture represents the error on the polarisation angle.}
\label{fig:fieldB_stars}
\end{figure}

\begin{figure}[t!]
\centering \includegraphics[width=\linewidth]{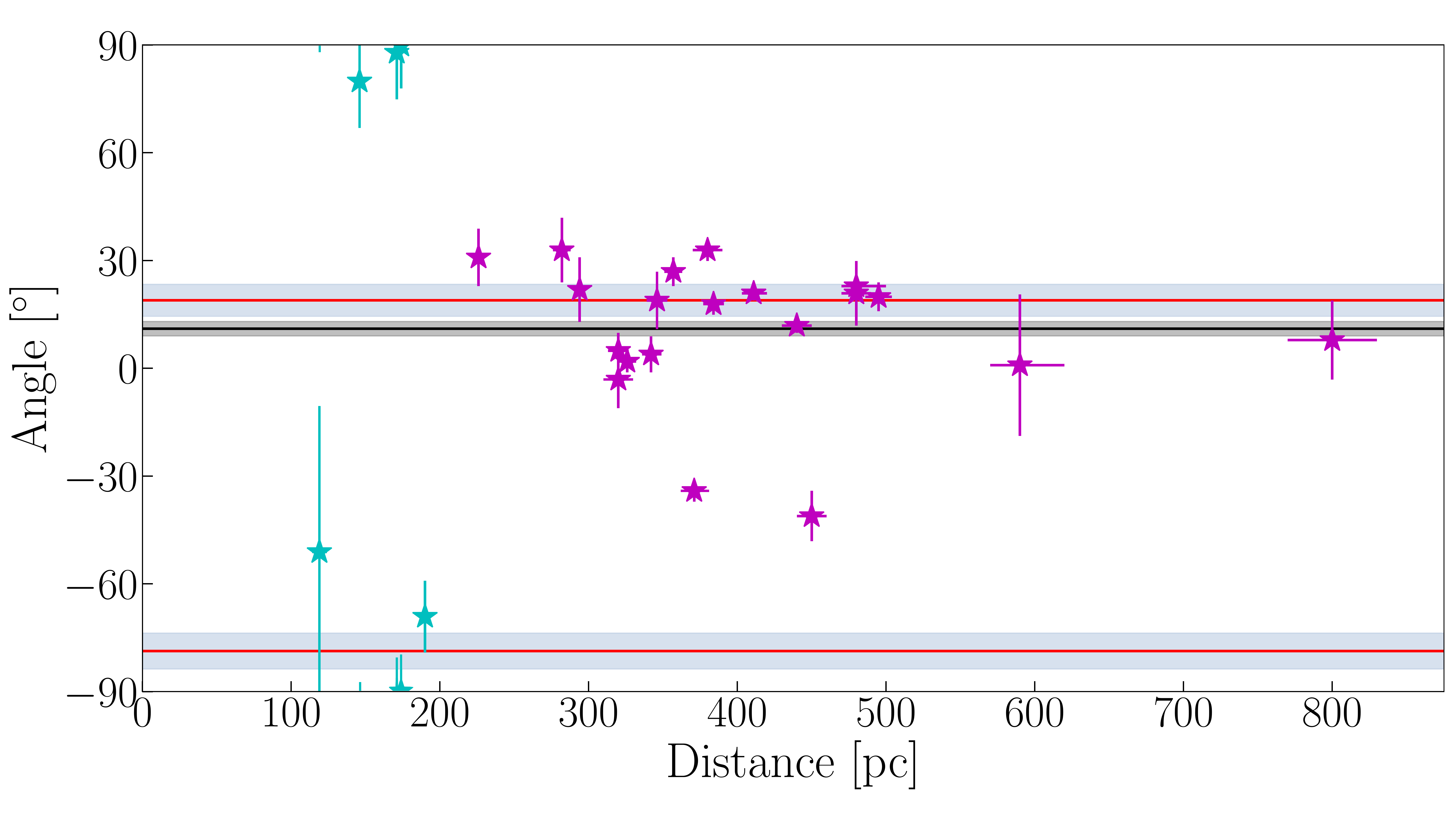}
\caption{Angles vs distance plot for Field B. The angle on the y-axis denotes the polarisation angle of the stars as well as the angle distribution of the depolarisation canals. The x-axis denotes the distance to the stars. Horizontal red lines mark the averages of distinctive orientations of the depolarisation canals, and light blue areas span the spread of the distributions for the corresponding averaged values, the same as in Figs.~\ref{fig:RHT_LOFAR_B} and \ref{fig:RHT_LOFAR_C} with two peaks $\sim$ $100^\circ$ apart. The colours of the stars correspond to the same groups of stars shown in Fig.~\ref{fig:fieldB_stars}. We can freely shift the polarisation angles of the stars modulo $\pi,$ meaning that the three stars plotted near $90^\circ$ could also belong to the area below $-90^\circ$ around the negative peak of the angle distribution of the depolarisation canals. The horizontal black line marks the average true value of the polarisation angle for the background stars, and the grey area spans the error range.}
\label{fig:lofar_star_polarization}
\end{figure}

\section{Starlight polarisation data - tracing distances}\label{sec:stars}
In this section, we present an innovative analysis that compares the orientation of the depolarisation canals with that of the polarisation angle detected towards stars in the fields of view. We searched for stars with available starlight polarisation data in the \texttt{VizieR}\footnote{\url{http://vizier.u-strasbg.fr}} database \citep{heiles00, berdyugin01, berdyugin&teerikorpi02, bailey&lucas&hough10, berdyugin&piirola&teerikorpi14}. We then extracted their distances from the Bailer-Jones catalogue \citep{bailer_jones18}, which is based on Gaia Data Release 2 \citep{gaiaDR2}.

In the case of Fields A and C, we unfortunately lack a large enough sample of starlight polarisation measurements to make a meaningful analysis. Thus, we only present results for Field B, where we find 25 stars with measured polarisation (see Table~\ref{tab:pol_stars_data} in Appendix~\ref{pol_stars_data}).
Their locations within the field are given in Fig.~\ref{fig:fieldB_stars}. Their measured polarisation angles versus their distances are plotted in Fig.~\ref{fig:lofar_star_polarization}. Error bars are plotted for both polarisation angles and distances. The error on the distance is larger if the star is farther away, ranging from a few parsecs for the close-by stars to a few dozen parsecs for the more distant stars. 
The polarisation angles of stars are grouped around two main orientations of the depolarisation canals presented in Figs.~\ref{fig:RHT_LOFAR_B} and \ref{fig:RHT_LOFAR_C}, that is, at $-80^\circ$ and $+20^\circ$, marked with horizontal lines in Fig.~\ref{fig:lofar_star_polarization}. The two stars with polarisation angles close to $+90^\circ$ are associated with the $-80^\circ$ orientation of the depolarisation canals due to the modulo $\pi$ convention in the polarisation angle. The distances are between 100 and 200 pc for stars with polarisation angle of about $-80^\circ$ and between 200 and 800 pc for stars with a polarisation of about $20^\circ$. Distance centroids are $161 \pm 1$ pc and $302 \pm 1$ pc for the first and second group of stars, respectively. 

The given starlight polarisation measurement directly traces the plane-of-the-sky magnetic field orientation, averaged over the dust density-weighted path-length to the star. Due to the complementary role of starlight polarisation produced by differential extinction in the visible and dust-polarised emission at longer wavelengths, the stars grouped at the largest distance likely trace the same average orientation of the magnetic field as is measured by the {\it Planck} satellite, namely the orientation of about $20^\circ$ with respect to the Galactic plane (see Sect.~\ref{ssec:Planck}). The group of stars in the foreground, in contrast, shows a typical magnetic field orientation of about $-80^\circ$, suggesting a tilt in the magnetic field structure between 200 and 300 pc.  

\begin{figure*}[h!]
\centering
\begin{subfigure}{0.3\textwidth}
\centering \includegraphics[width=\linewidth]{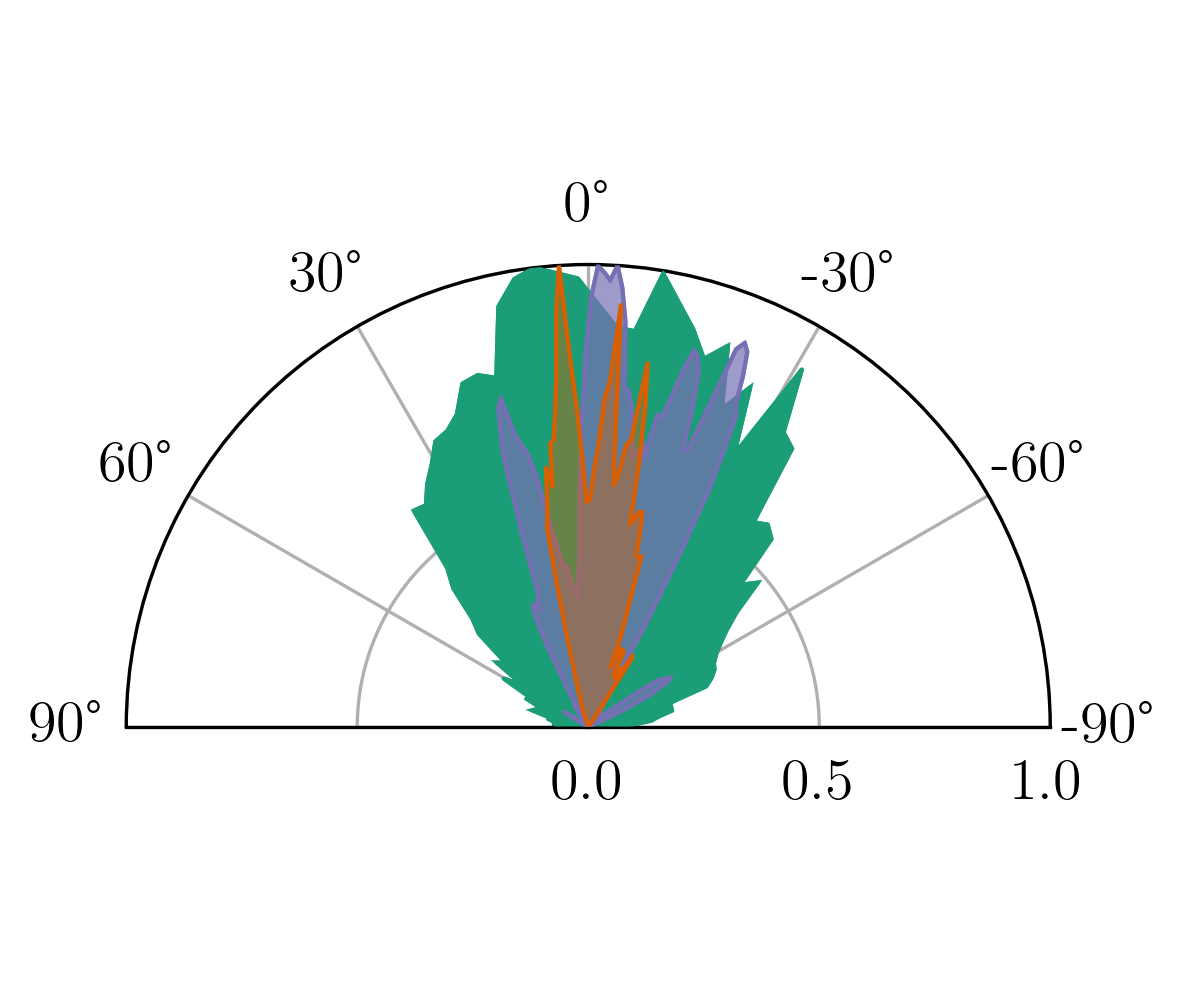}
\caption{Field A}
\label{fig:LOFAR_EBHIS_Planck_A}
\end{subfigure}
\begin{subfigure}{0.3\textwidth}
\centering \includegraphics[width=\linewidth]{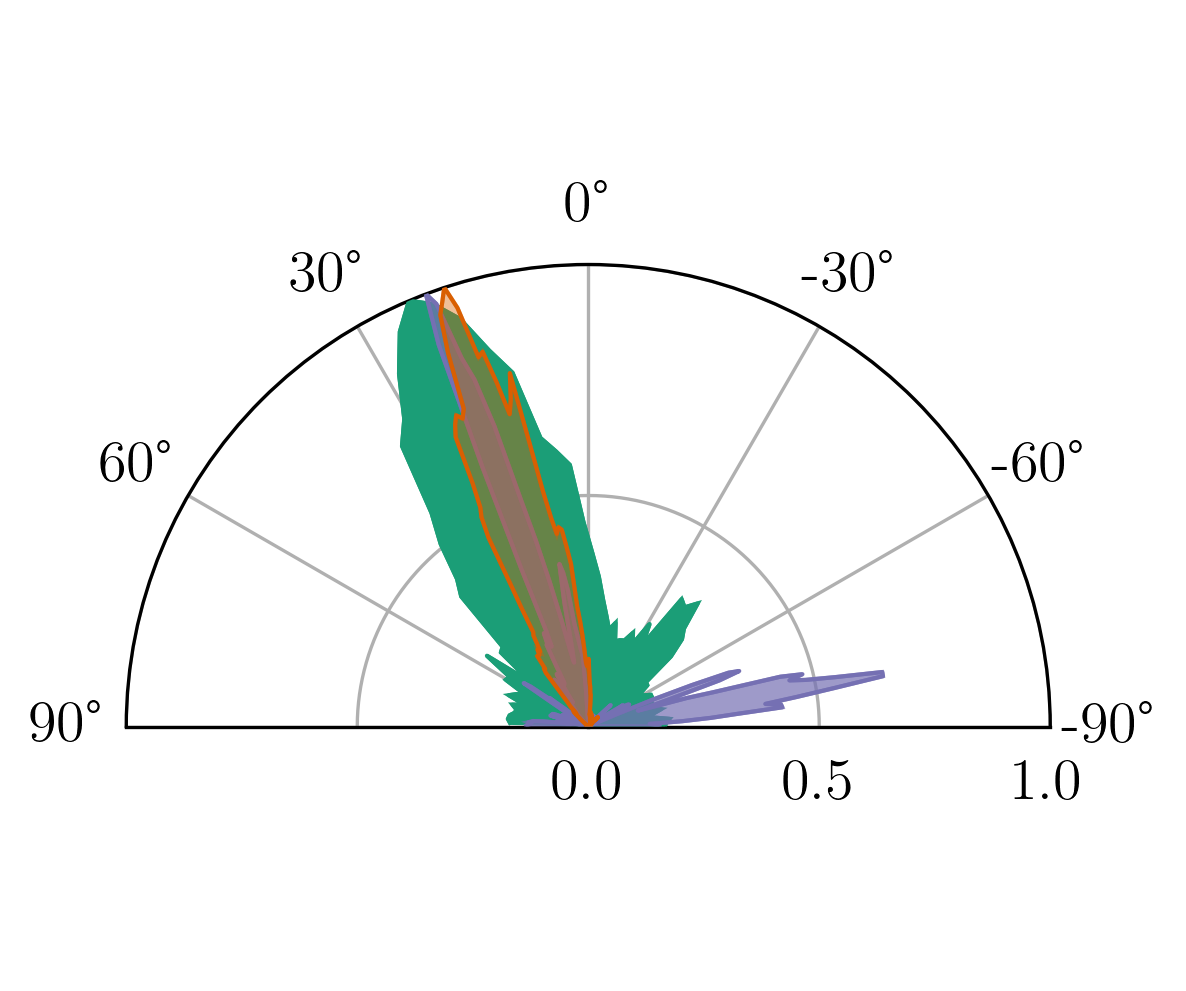}
\caption{Field B}
\label{fig:LOFAR_EBHIS_Planck_B}
\end{subfigure}
\begin{subfigure}{0.3\textwidth}
\centering \includegraphics[width=\linewidth]{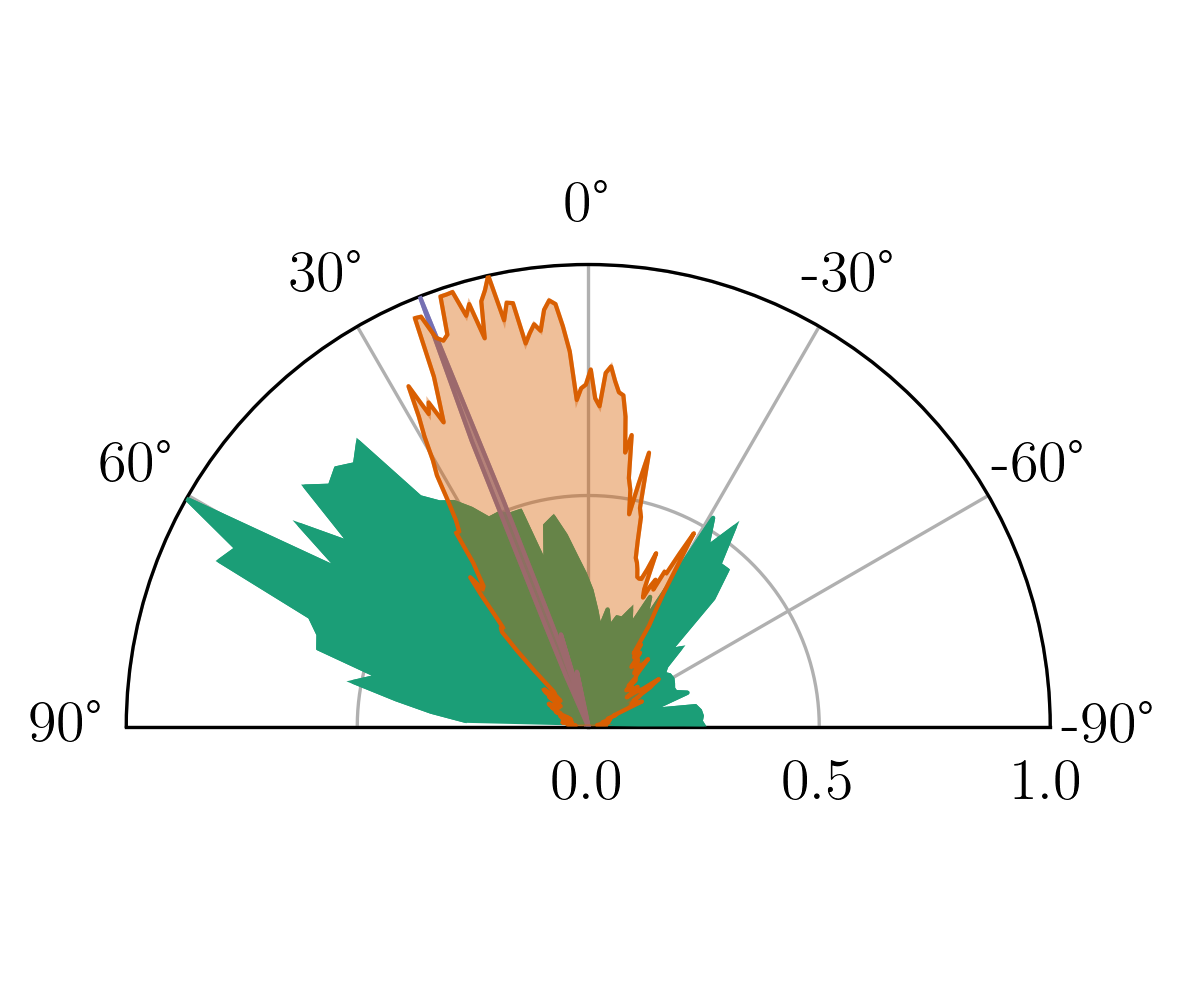}
\caption{Field C}
\label{fig:LOFAR_EBHIS_Planck_C}
\end{subfigure}
\caption{Summary of the results (see Fig.~\ref{fig:RHT_LOFAR_A} -- \ref{fig:Planck_ABC}) showing the comparison of the orientations of the depolarisation canals (purple distribution), the $\text{H}\textsc{I}$ filaments (green distribution), and the plane-of-the-sky magnetic field component (orange distribution) in relative scale. For Field A, we summed RHT results for depolarisation canals identified across the three Faraday-depth ranges given in Fig.~\ref{fig:RHT_LOFAR_A}.}
\label{fig:LOFAR_EBHIS_Planck_ABC}
\end{figure*}

We note, however, that the presence of these two separate groups of stars along the sight line with different magnetic field orientations requires a more careful discussion of the comparison between the orientation of the depolarisation canals and that of the starlight polarisation. As a typical problem of magnetic field tomography based on starlight polarisation data \citep[i.e.][]{Panopoulou19}, we need to de-rotate the observed polarisation of the stars in the background for the amount of polarisation detected in the foreground. As explained in Appendix~\ref{pol_stars_data}, from the catalogue, we estimate the average Stokes $q$ and $u$ parameters (see Eq.~\ref{eq:stokes}) with corresponding errors for the two distinct groups of stars, ($\bar{q}_{\rm near},\bar{u}_{\rm near}$)\footnote{The bar refers to average quantities.} and ($\bar{q}_{\rm far},\bar{u}_{\rm far}$). The average true value of the polarisation angle for the background stars can be written as $\bar{\psi}^{\rm true}_{\rm far} = 0.5 \arctan{(\bar{u}^{\rm true}_{\rm far},\bar{q}^{\rm true}_{\rm far})}$, where $\bar{u}^{\rm true}_{\rm far} = \bar{u}_{\rm far} - \bar{u}_{\rm near}$ and $\bar{q}^{\rm true}_{\rm far} = \bar{q}_{\rm far} - \bar{q}_{\rm near}$. Based on the tabulated data, we obtain a value of $\bar{\psi}^{\rm true}_{\rm far}= 11^{\circ} \pm 2^{\circ}$, which is still rather consistent with the orientation of the depolarisation canals shown in Figs.~\ref{fig:RHT_LOFAR_B} and \ref{fig:RHT_LOFAR_C}. The foreground stars do not seem to affect the estimate of the background starlight polarisation much probably because the ratio is low, roughly $\sim$30\%,  between the degree of polarisation of the nearby stars and that of the far away stars.

\section{Summary and conclusions}
We have presented results based on the LOFAR HBA observations of three fields (Fields A, B, and C) in the surroundings of the 3C 196 field. We used Faraday tomography to analyse the observed polarised emission, which appears at Faraday depths between $-10~{\rm rad~m^{-2}}$ and $+15~{\rm rad~m^{-2}}$. The polarised emission is brightest in Field A ($\sim$$5~{\rm K}$) because it is closest to the Galactic plane of the three fields. While the typical emission brightness in Field B is $\sim$$1.5~{\rm K}$, in Field C it reaches only $\sim$$0.6~{\rm K}$ as this field is located in the coldest part of the northern Galactic hemisphere. 

The observed morphology of emission differs in each field. It is richest in Field A, where we observe three distinct groups of structures at different Faraday depths. The morphology in Field B is dominated by a cross-like structure, while in Field C it is very faint, patchy, and diffuse. While all three fields show a complex system of straight depolarisation canals, they are more prominent in Fields A and B than in Field C.  A likely explanation of this is the fact that we observe a lower amount of emission in Field C than in the other two fields. These depolarisation canals are associated with the effect of beam depolarisation in regions of abrupt changes in polarisation angle. The depolarisation canals we observe extend to a few degrees in length and resemble those observed in field 3C 196 \citep{jelic15}. 

In Fields A, B, and C, \citet{bracco20} studied the statistical correlation between the polarised emission detected by LOFAR and the brightness temperature of the $\text{H}\textsc{I}$ emission at 21 cm. They found a strong correlation between the two tracers of the multiphase ISM in Field A, in particular, with CNM $\text{H}\textsc{I}$ gas, while they observed a lack of statistical correlation in Fields B and C. As explained by the authors, this resulted from limitations of the chosen statistical metrics given the patchy morphological emission in Fields B and C, as described above. 

We have focussed on the characteristics of the depolarisation canals. We compared their orientation with that of both $\text{H}\textsc{I}$ filaments and the plane-of-the-sky magnetic field component using the RHT.
The results of this analysis are summarised in Fig.~\ref{fig:LOFAR_EBHIS_Planck_ABC}. There is a clear alignment between the three distinct ISM tracers in Fields A and B. The observed orientation in Field A is very similar to the one observed in the 3C 196 field \citep[see Fig.~4 in][]{jelic18}. The dominant orientation of the depolarisation canals in Field B is $\sim15^\circ$ more inclined with respect to the Galactic plane than in Field A and in the 3C 196 field. The alignment found for the $\text{H}\textsc{I}$ filaments selected by the RHT is mostly associated with CNM structures \citep{clark14} aligned to the depolarisation canals. This supports the result reported by \citet{bracco20} in the same fields of view, but considering polarised intensity. 

Furthermore, in Field B, we observed two groups of stars at distances below and above 200 pc, respectively, that probe distinct magnetic field orientations. These are both comparable with the orientations of the depolarisation canals in the same field. The depolarisation canals seem to trace the same change of the magnetic field as probed by the stars. This change occurs at $\sim200~{\rm pc}$, a distance that is compatible with the edge of the Local Bubble, as also supported by 3D maps of the (local) ISM from measurements of starlight extinction by interstellar dust \citep[][STILISM\footnote{https://stilism.obspm.fr/} data]{lallement14, capitanio17, Lallement19}, where the extinction seems to increase mostly between 150 and 300 pc.

Because starlight polarisation data are currently limited, however, it is difficult to assess how statistically significant this result is. Future stellar polarisation surveys of the high Galactic latitude sky, such as the Polar-Area Stellar Imaging in Polarisation High-Accuracy Experiment \citep[PASIPHAE; ][]{Tassis18}, will change this and allow us to make a more detailed analysis. 

 The change in magnetic field traced by the stars and the depolarisation canals is also seen with the orientations of the $\text{H}\textsc{I}$ filaments presented in this work. The $\text{H}\textsc{I}$ filaments show an alignment over several velocity channels and a coherent change in their orientations towards the highest velocities. As discussed by \citet{clark18} and \citet{clark19}, this behaviour is expected if the magnetic field is coherent, mostly in the plane of the sky, but twisted along the line of sight. This is supported by the higher value of the synthetic $\text{H}\textsc{I}$ polarisation fraction ($p_{\text{H}\textsc{I}}$) in this field, calculated from the velocity-integrated synthetic $\text{H}\textsc{I}$ Stokes parameters maps \citep{clark19}.  

Field A also shows the higher $p_{\text{H}\textsc{I}}$ than the all-sky average, as is reported for the 3C 196 field  \citep{clark19}. Therefore the observed alignment between the ISM tracers reported in this work (Fields A and B) and in the 3C 196 field \citep{zaroubi15, jelic18, bracco20} indicates a common, very ordered magnetic field with a dominant component in the plane of the sky, which probably shapes the observed morphology in a wide-field area of the sky connecting these three fields. 

In contrast, this is not the case in Field C. We do not observe a clear correlation between the ISM tracers there, $p_{\text{H}\textsc{I}}$ is smaller than the all-sky average, and the plane of the sky magnetic field component probed by the Planck is less ordered. Hence, the dominant magnetic field component in Field C is probably not so much in the plane of the sky, and it is more random than ordered. 

To conclude, multi-tracer analyses of Faraday tomographic data are inevitable if distances to the observed structures are to be constrained and the 3D nature of the magnetic field is to be understood. The magnetic field needs to be ordered with a dominant component in the plane of the sky to observe a correlation between different tracers of the multi-phase ISM. By combining Faraday tomographic data with starlight polarisation data we were able for the first time to directly estimate the distance to the observed depolarisation canals in the LOFAR data. The straight depolarisation canals seem to change orientation at the edge of the Local Bubble. 

\begin{acknowledgements}
We thank an anonymous referee for their valuable comments which improved the paper. LT, VJ and AE acknowledge support by the Croatian Science Foundation for a project IP-2018-01-2889 (LowFreqCRO) and additionally LT and VJ for the project DOK-2018-09-9169. MH acknowledges funding from the European Research Council (ERC) under the European Union's Horizon 2020 research and innovation programme (grant agreement No. 772663). AB acknowledges the support from the European Union’s Horizon 2020 research and innovation program under the Marie Sk{\l}odowska-Curie Grant agreement No. 843008 (MUSICA). This paper is based on data obtained with the International LOFAR Telescope (ILT) under project code LC5\_008. LOFAR \citep{haarlem13} is the Low Frequency Array designed and constructed by ASTRON. It has observing, data processing, and data storage facilities in several countries, that are owned by various parties (each with their own funding sources), and that are collectively operated by the ILT foundation under a joint scientific policy. The ILT resources have benefited from the following recent major funding sources: CNRS-INSU, Observatoire de Paris and Universit\'e d’Orl\'eans, France; BMBF, MIWF-NRW, MPG, Germany; Science Foundation Ireland (SFI), Department of Business, Enterprise and Innovation (DBEI), Ireland; NWO, The Netherlands; The Science and Technology Facilities Council, UK; Ministry of Science and Higher Education, Poland. The processing of the LOFAR observations were done on a CPU/GPU cluster dedicated to the LOFAR-EoR project, located at the University of Groningen and ASTRON in the Netherlands. Some of the results in this paper have been derived using the healpy and HEALPix package.
\end{acknowledgements}

\bibliographystyle{aa}
\bibliography{3C196area.bbl}
\onecolumn{
\begin{appendix}
\section{Starlight polarisation data}
\label{pol_stars_data}

This table gives the coordinates, distance, and polarisation angle of the 25 stars we used in the analysis of Field B (Sect.~\ref{sec:stars}). The data are based on a selection of stars from the polarisation catalogues \citep{heiles00, berdyugin01, berdyugin&teerikorpi02, bailey&lucas&hough10, berdyugin&piirola&teerikorpi14}. Their distances are extracted from the Bailer-Jones catalogue \citep{bailer_jones18}, which is based on Gaia Data Release 2 \citep{gaiaDR2}. The polarisation angles are given with respect to the Galactic plane in the range between $-90^{\circ}$ and $+90^{\circ}$, consistently with the analysis of the depolarisation canals, the $\text{H}\textsc{I}$ filaments, and the plane-of-the-sky magnetic field. The measurements of starlight polarisation allowed us to retrieve the Stokes parameters (Stokes $Q$ and $U$) relative to the total intensity (Stokes $I$) for each star, which we estimated from the degree of polarisation, $p$, and the polarisation angle, $\psi$, as follows:
\begin{equation}\label{eq:stokes}
    q = Q/I = p \cos{2\psi} \,\hspace{3mm}; \hspace{3mm}
    \,u = U/I = p \sin{2\psi}.
\end{equation}

\begin{table}[ht!]
\caption{List of 25 stars in Field B we used in the analysis, with their positions, distances, and polarisation angles (sorted by distance). A dashed horizontal line separates the two groups of stars discussed in Sect.~\ref{sec:stars}.}       
\label{tab:pol_stars_data}      
\centering          
\begin{tabular}{ c | c | c | c}    
\hline\hline       
   Position (RA, Dec) [$^\circ$] & Distance $\mathrm{[pc]}$ & Polarisation angle $\psi$ [$^\circ$] & Degree of polarisation $p \ [\rm \%]$ \\  \hline 
   125.708664, +43.187697 & $119^{+5}_{-5}$ & $-51.1 \pm 40.6$ & $0.070 \pm 0.120$ \\ 
   120.447527, +40.144664 & $146^{+1}_{-1}$ & $79.9 \pm 13.0$ & $0.037 \pm 0.017$ \\
   120.303935, +44.805100 & $171^{+1}_{-1}$ & $87.9 \pm 13.0$ & $0.074 \pm 0.036$ \\
   123.725076, +40.578497 & $174^{+2}_{-2}$ & $89.9 \pm 12.0$ & $0.048 \pm 0.020$ \\    
   127.833110, +37.264814 & $190^{+4}_{-3}$ & $-69.1 \pm 10.0$ & $0.105 \pm 0.038$ \\ 
   \hdashline
   125.644908, +45.350866 & $226^{+3}_{-2}$ & $30.9 \pm 8.0$ & $0.110 \pm 0.031$ \\
   121.355194, +41.190057 & $282^{+6}_{-6}$ & $32.9 \pm 9.0$ & $0.162 \pm 0.053$ \\     
   128.182245, +42.139957 & $294^{+4}_{-4}$ & $21.9 \pm 9.0$ & $0.109 \pm 0.034$ \\     
   125.141953, +38.345584 & $320^{+10}_{-10}$ & $-3.1 \pm 8.0$ & $0.090 \pm 0.026$ \\
   129.168437, +39.418853 & $320^{+7}_{-7}$ & $4.9 \pm 5.0$ & $0.170 \pm 0.029$ \\     
   126.509845, +36.982582 & $326^{+6}_{-6}$ & $1.9 \pm 3.0$ & $0.211 \pm 0.022$ \\     
   123.987250, +44.277259 & $342^{+7}_{-6}$ & $3.9 \pm 5.0$ & $0.164 \pm 0.030$ \\     
   123.952519, +35.939286 & $346^{+7}_{-7}$ & $18.9 \pm 8.0$ & $0.110 \pm 0.030$ \\     
   126.696234, +39.892149 & $357^{+6}_{-6}$ & $26.9 \pm 4.0$ & $0.193 \pm 0.026$ \\     
   127.768099, +35.362672 & $371^{+10}_{-9}$ & $-34.1 \pm 3.0$ & $0.295 \pm 0.033$ \\
   125.434397, +43.333974 & $380^{+10}_{-10}$ & $32.9 \pm 3.0$ & $0.490 \pm 0.043$ \\
   127.402977, +42.485128 & $384^{+7}_{-7}$ & $17.9 \pm 3.0$ & $0.231 \pm 0.028$ \\     
   122.205098, +37.226994 & $411^{+9}_{-8}$ & $20.9 \pm 2.0$ & $0.213 \pm 0.017$ \\     
   121.951794, +36.243770 & $440^{+10}_{-10}$ & $11.9 \pm 2.0$ & $0.259 \pm 0.021$ \\
   128.401712, +35.798222 & $450^{+10}_{-10}$ & $-41.1 \pm 7.0$ & $0.089 \pm 0.023$ \\
   129.961735, +41.365977 & $480^{+10}_{-10}$ & $20.9 \pm 9.0$ & $0.152 \pm 0.047$ \\
   125.909633, +40.674991 & $480^{+20}_{-10}$ & $22.9 \pm 3.0$ & $0.214 \pm 0.024$ \\    
   128.186731, +43.624892 & $495^{+9}_{-9}$ & $19.9 \pm 4.0$ & $0.168 \pm 0.025$ \\     
   119.645290, +43.503888 & $590^{+30}_{-20}$ & $0.9 \pm 19.7$ & $0.280 \pm 0.200$ \\ 
   129.873905, +44.291835 & $800^{+30}_{-30}$ & $7.9 \pm 11.0$ & $0.119 \pm 0.048 $ \\ 
   \hline
\end{tabular}
\end{table}
\end{appendix}}
\end{document}